\documentclass[onecolumn,11pt]{article}
\usepackage[top=1in, bottom=1in, left=1in, right=1in]{geometry}
\usepackage{bm,amsmath,amsfonts,amscd,amssymb}
\usepackage{graphicx}
\usepackage{url}
\usepackage{caption}
\usepackage{color}
\usepackage{mathtools}
\usepackage{setspace}
\usepackage[numbers,sort&compress]{natbib}
\usepackage{framed}
\usepackage{enumitem}
\usepackage{newtxtext}
\usepackage{newtxmath}
\usepackage{hyperref}
\hypersetup{
  colorlinks = true,
  urlcolor  = blue,
  citecolor = black,
}
\usepackage{fancyhdr}

\newcommand{\RomanNumeralCaps}[1]
\linenumbers

\newcommand\blfootnote[1]{%
  \begingroup
  \renewcommand\thefootnote{}\footnote{#1}%
  \addtocounter{footnote}{-1}%
  \endgroup
}

\makeatletter
\renewcommand\paragraph{\@startsection{paragraph}{4}{\z@}%
                                     {-3.25ex\@plus -1ex \@minus -.2ex}%
                                     {0.0001pt \@plus .2ex}%
                                     {\normalfont\normalsize\bfseries}}
\renewcommand\subparagraph{\@startsection{subparagraph}{5}{\z@}%
                                     {-3.25ex\@plus -1ex \@minus -.2ex}%
                                     {0.0001pt \@plus .2ex}%
                                     {\normalfont\normalsize\bfseries}}

\counterwithin{paragraph}{subsubsection}
\counterwithin{subparagraph}{paragraph}
\makeatother

\title{\LARGE{\vspace{-.55in}\textbf{Cluster-based decomposition}}\vspace{-.175in}}
\title{\vspace{-.55in}{\fontsize{16}{16}\selectfont \textbf{Cluster regression model for control of nonlinear dynamics}}\vspace{-.15in}}

\author{\normalsize{Nitish Arya$^{1*}$, Aditya G.~Nair$^{1}$}\\
\footnotesize{$^1$ Department of Mechanical Engineering, University of Nevada, Reno, NV 89557}\\
}
\date{}

\begin{document}
\maketitle

\blfootnote{$^*$ Corresponding author (adityan@unr.edu).}
\vspace{-.2in}

\begin{abstract}

In the realm of big data, discerning patterns in nonlinear systems affected by external control inputs is increasingly challenging. Our approach blends the coarse-graining strengths of centroid-based unsupervised clustering with the clarity of sparse regression in a unique way to enhance the closed-loop feedback control of nonlinear dynamical systems. A key innovation in our methodology is the employment of cluster coefficients via a cluster decomposition of time-series measurement data. This approach transcends the conventional emphasis on the proximity of time series measurements to cluster centroids, offering a more nuanced representation of the dynamics within phase space. Capturing the evolving dynamics of these coefficients enable the construction of a robust, deterministic model for the observed states of the system. This model excels in capturing a wide range of dynamics, including periodic and chaotic behaviors, under the influence of external control inputs. Demonstrated in both the low-dimensional Lorenz system and the high-dimensional scenario of a flexible plate immersed in fluid flow, our model showcases its ability to pinpoint critical system features and its adaptability in reaching any observed state. A distinctive feature of our control strategy is the novel hopping technique between cluster states, which successfully averts lobe switching in the Lorenz system and accelerates vortex shedding in fluid-structure interaction systems while maintaining the mean aerodynamic characteristics. 

\end{abstract}

\section{Introduction}

In the realm of scientific and engineering disciplines, the control of dynamical systems is a critical endeavor, now more challenging in the age of big data. The high-dimensionality characteristic of many physical systems, compounded by the complexities introduced by external control inputs, demands innovative approaches for effective management. Recent advancements have focused on distilling these complex systems into manageable models, employing both linear~\cite{rowley2009spectral, schmid2010dynamic, herve2012physics} and nonlinear regression techniques~\cite{brunton2016discovering, billings2013nonlinear}, with a specific emphasis on integrating exogenous forces \cite{proctor2016dynamic, brunton2016sparse, kaiser2018sparse}.

Central to this evolution is the understanding that data, particularly from large datasets, tend to align with a low-dimensional manifold, suggesting an inherent structure within the apparent complexity \cite{fefferman2016testing}. This has led to the development of various modeling techniques, from linear methods like Proper Orthogonal Decomposition (POD) \cite{berkooz1993proper, holmes2012turbulence} to more nuanced nonlinear methods such as locally linear embedding \cite{donoho2003hessian, roweis2000nonlinear}. These approaches are pivotal in crafting reduced order models (ROMs) that can predict system dynamics over time, as seen in the success of the POD-Galerkin model \cite{rowley2017model, noack2011reduced} and more recent innovations using deep autoencoders \cite{lusch2018deep, champion2019data}, neural networks \cite{floryan2022data}, and spectral submanifolds \cite{cenedese2022data}.

Despite these advancements, modeling system dynamics accurately under variable control conditions remains a formidable challenge \cite{brunton2015closed, benner2015survey}. The introduction of external control variables often propels systems into unexplored regions of the state space, where traditional control methods, when applied to ROMs, require extensive calibration, particularly in nonlinear and disturbed environments \cite{callaham2022role, zucatti2021calibration, wan2018data}. This highlights the need for flexible and robust control strategies. A more comprehensive technical background for this study is detailed in Appendix \ref{appendixA}.

Our research introduces an innovative cluster-based method that offers a fresh perspective on system representation. By harnessing data similarities, this method effectively reduces the complexity of analyzing vast datasets to studying key representative clusters, a technique finding utility across various domains including engineering and misinformation detection \cite{murphy2012machine, hosseinimotlagh2018unsupervised}. This approach also enriches turbulent flow analysis and control mechanisms \cite{kaiser2014cluster, nair2019cluster}, combining the clarity of data-driven models with the intuitiveness of traditional dynamical modeling. Yet, these models often assume Markovian dynamics, which may limit their predictive capabilities and raise questions about the assumption of temporal independence (see Appendix \ref{appendixB2}). Probabilistic models, while offering in-depth insights, tend to lose predictive accuracy over time (see Appendix \ref{appendixC1}).

Addressing these limitations, our approach introduces an innovative application of cluster coefficients, derived through a novel decomposition process of time-series data. This methodology advances beyond the conventional reliance on proximity metrics for cluster centroids, offering a more intricate depiction of the system's dynamics in the phase space. This advancement paves the way for the development of a resilient and deterministic model, proficient in capturing an extensive array of dynamic behaviors, from the regularity of periodic patterns to the unpredictability of chaotic motions, particularly in response to external stimuli. The versatility and accuracy of our model are demonstrated through its application in diverse settings: it proves effective in simpler, low-dimensional systems such as the Lorenz system, as well as in more complex, high-dimensional scenarios like the fluid dynamics of a compliant flat plate. These examples highlight the model's remarkable ability to assimilate knowledge from a limited amount of training data and its precision in steering systems toward specific desired outcomes. Additionally, a novel and notable aspect of our control technique is the 'hopping' strategy between cluster states. This strategy is particularly effective in preventing lobe switching in systems like the Lorenz model and promotes enhanced vortex shedding in fluid-structure interaction scenarios, all while preserving the essential aerodynamic properties of the system.

This paper presents a significant advancement in the control of nonlinear dynamical systems, especially pertinent in the era of big data. By integrating centroid-based clustering with sparse regression within a cluster regression framework, we provide a powerful tool for managing complex systems, marking a significant stride in dynamic system control and analysis.

 \begin{figure}[p]
    \centering
    \includegraphics[width=0.85\textwidth]{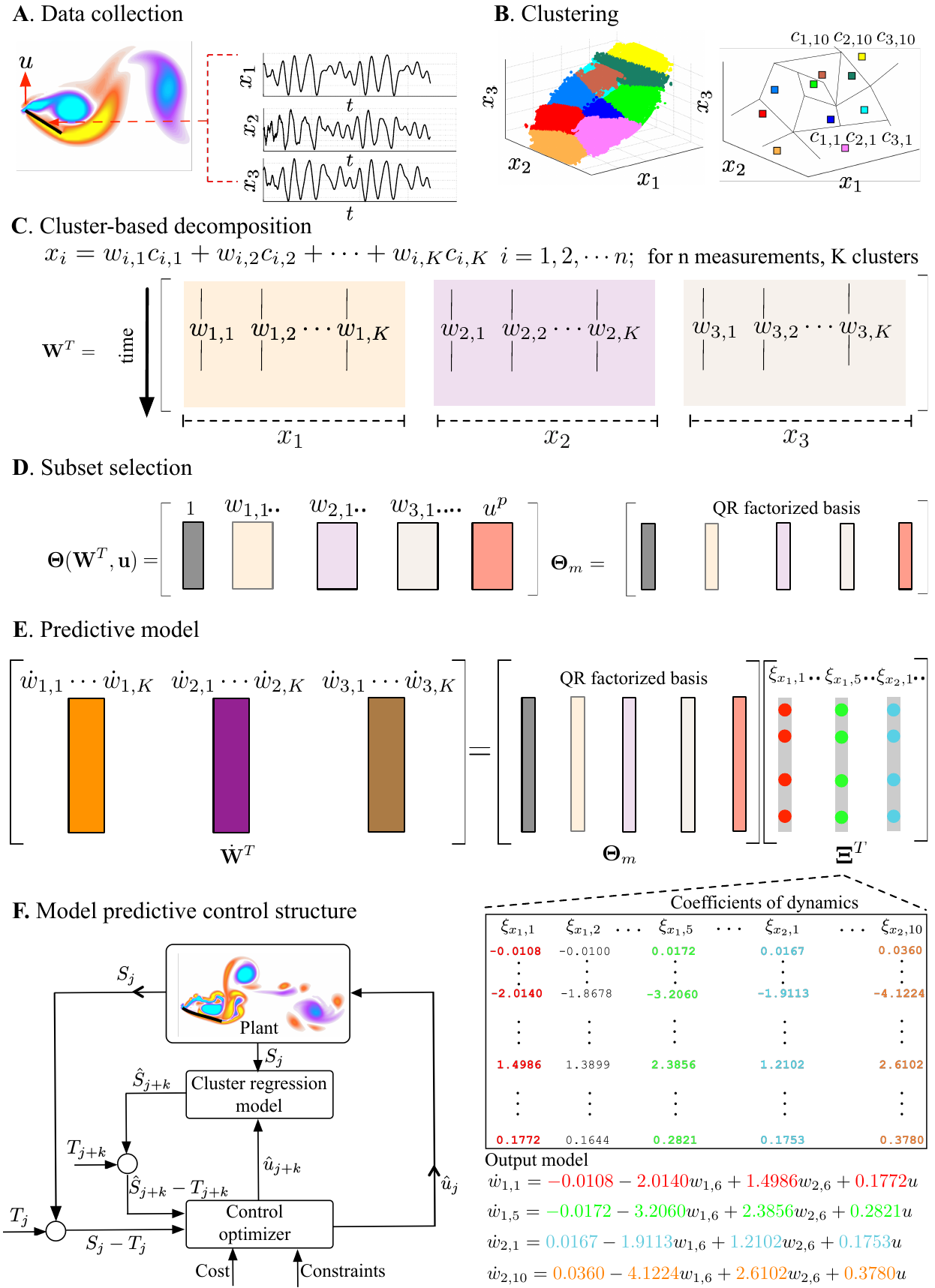}
    \caption{Overview of the cluster regression model for control of a flow over a compliant flat plate at an angle of attack. \textbf{A}. Collection of limited sensor measurements from experiments or simulations, such as lift and drag coefficients, using Direct Numerical Simulation (DNS). \textbf{B}. Application of k-means clustering to these measurements to identify $K$ distinct centroids. \textbf{C}. Decomposition of the measurements into cluster coefficients based on the identified centroids. \textbf{D}. Selection of an optimal library through QR decomposition with column pivoting for enhanced model accuracy and reducing computational complexity.  \textbf{E}. Construction of a predictive model for the coefficients employing polynomial regression of order $p$ for robust prediction.  
    \textbf{F}. Deployment of the coefficient model in trajectory prediction, guiding the system towards desired cluster states via model predictive control for optimal performance.}
    \label{fig:1}
\end{figure}

\section{Cluster regression framework}

Our cluster regression framework comprises of three main stages as outlined below. 

\vspace{-0.1in}
\subsection{Cluster-based decomposition}

The development of our cluster regression model commences with the collection of $n$ time series measurements $\mathbf{x} = [x_1~x_2~\cdots~x_n]$ from a dynamical system under the influence of an external force $\mathbf{u} = [{u}_1, {u}_2, \ldots, {u}_n]$, each corresponding to a dimension in the feature space (refer to Fig. \ref{fig:1}(A) for visual representation). This model leverages the k-means clustering algorithm to partition these time-series measurements into $K$ distinct clusters (illustrated in Fig. \ref{fig:1}(B)), where each measurement is grouped based on proximity to the nearest cluster centroid \cite{hartigan1979algorithm}. The k-means method aims to create \(K\) separate, non-overlapping subsets, minimizing the within-cluster variance. This variance is defined as the cumulative sum of squared distances between each data point and its cluster's centroid. The process involves iterative recalculations of the centroids, continuing until the cluster assignments stabilize or certain predefined criteria, like a specific tolerance level or a maximum number of iterations, are met. To ensure robustness in clustering outcomes, multiple runs with varying initial conditions are typically conducted \cite{arthur2007k}. Further details on the clustering methodology can be found in Appendix \ref{appendixB1}.

In our model, each measurement $x_i$ at time $t$ is affected by system parameters $\theta$, which, in high-dimensional contexts, may correspond to spatial points. Drawing upon the principles of modal decomposition \cite{taira2017modal}, a measurement can be broken down into a combination of basis vectors and their corresponding temporal coefficients. Consequently, we can represent each measurement using the following equation:

\begin{equation} 
\label{eq:8}
x_i(t,\theta) = f(w_{i,k}(t), c_{i,k}(\theta)) = \sum_{k=1}^{K} w_{i,k}(t)c_{i,k}(\theta), \quad i=1,2, \dots, n
\end{equation}
where the coefficients, $w_{i,k}$, function analogously to the temporal coefficients found in Proper Orthogonal Decomposition (POD) and are referred to as cluster coefficients in our study. Here, $c_{i,k}$ represents the cluster centroids. The first subscript for the cluster coefficients and centroids indicates the feature space coordinate and the second subscript indicates the cluster number. We can also extend this decomposition to the nonlinear combination of cluster centroids and associated weight coefficients, akin to auto-encoder representation in machine learning \cite{champion2019data}. In the probabilistic model, $w_{i,k}$ represents the cluster state probability (see Appendix \ref{appendixB2}).

In general, the cluster coefficients $w_{i,k}$ could be determined through various strategies. For instance, if employing inverse distance weighting, the coefficients would be inversely proportional to the distance between data points, emphasizing nearer neighbors more significantly. Alternatively, using a $k$-nearest neighbors approach, the coefficients would be based on the proximity to the $k$ closest centroids, providing a localized perspective on the data's structure. In the subsequent stage of our methodology, we determine the cluster coefficients by employing a sparse regression approach. This technique is pivotal in identifying the most relevant and significant coefficients, thereby streamlining the model by focusing on the most impactful data features. 

\subsection{Predictive model}

Continuing from the decomposition phase, we proceed to derive a predictive model for the cluster coefficients utilizing nonlinear sparse regression \cite{brunton2016discovering,kaiser2018sparse}. This method is crucial for capturing the non-linear dynamics inherent in complex systems. A key aspect of this process involves training the model with appropriate control inputs, integrating both the coefficients and inputs to accurately reflect the system's behavior. To enhance computational efficiency and manageability, especially important in high-dimensional settings, we consolidate all the derived cluster coefficients into a single matrix:
\begin{equation}
\mathbf{W} = [{w}_{1,1} \cdots {w}_{1,K} \hspace{0.5cm} {w}_{2,1} \cdots {w}_{2,K} \hspace{0.2cm} \cdots \hspace{0.2cm} {w}_{n,1} \cdots {w}_{n,K}]^T 
\end{equation}
as shown in Fig. \ref{fig:1}(C).

Let's consider a non-linear dynamical system represented as
\begin{equation} 
\dot{\mathbf{W}} = \mathbf{F}(\mathbf{W},\mathbf{u}),
\label{eq:reg}
\end{equation}
where $\mathbf{u} = [{u}_1, {u}_2, \ldots, {u}_m]$ symbolizes the $m$ external control inputs. For the purposes of the present work, only a single input forcing parameter is considered. To unravel the system's intrinsic dynamics, we develop a comprehensive library of polynomial functions, denoted as $\Theta(\mathbf{W},\mathbf{u})$, which extends up to a specified order $p$. This library is constructed from the cluster coefficients in tandem with the input forcing. Consequently, $\Theta(\mathbf{W},\mathbf{u})$ consists of $\binom{n \times K+p+1}{p}$ columns, encompassing a wide array of potential interactions and nonlinear effects within the system. When the time derivatives of feature space variables are available, they can be recast as derivatives of cluster coefficients; in their absence, numerical differentiation is applied. We can rewrite Eq.~\ref{eq:reg} using the library of polynomial functions as
\begin{equation}
\dot{\mathbf{W}} = \mathbf{\Xi} \mathbf{\Theta}^{T}(\mathbf{W},\mathbf{u}).
\label{eq:4}
\end{equation} 
where $\Xi$ are the regression coefficients. Sparse regression, despite its computational efficiency for smaller datasets, can incur significant overhead when handling even a limited set of measurements in a feature space accompanied by a moderate cluster count. As an illustrative example, consider the Lorenz system, detailed further in subsequent sections. This system has a three-dimensional feature space ($n = 3$) segmented into ten clusters ($K = 10$), and it employs a polynomial order $p = 2$ for the library terms. As a result, the library encompasses a vast 528 columns. Beyond the computational implications, this expansive library size can lead to numerous prospective models. 

To select the optimal model, one needs to balance between its complexity and accuracy. Such a balance can be achieved using statistical benchmarks like the Bayesian information criterion (BIC) \cite{schwarz1978estimating} or the Akaike information criterion (AIC) \cite{akaike1998information,akaike1974new}, as demonstrated in \cite{mangan2017model}. Our approach harnesses the insights from cluster-based decomposition to prune the library functions by identifying all the linearly independent columns (or threshold dominant columns) of $\mathbf{\Theta}(\mathbf{W},\mathbf{u})$ using QR factorization with column pivoting \cite{van1996matrix,miller2002subset} (as visualized in Fig. \ref{fig:1} (D)). The refined library, $\mathbf{\Theta}_m$ is characterized by fewer columns than its predecessor, $\mathbf{\Theta}$. Specifically, the library column count for the Lorenz system is efficiently reduced to $15$. The subsequent cluster regression model is given by
\begin{equation}
\dot{\mathbf{W}} = \mathbf{\Xi} \mathbf{\Theta}_m.
\label{eq:6}
\end{equation} 
The coefficients $\mathbf{\Xi}$ are obtained from the minimization procedure as
\begin{equation}
\xi_j = \underset{\hat{\xi_j}} {\textup{argmin}} \frac{1}{2} \left\lVert \dot{W}_j - \hat{\xi_j} \Theta_m^{T}\right\rVert_2^2 + \lambda \left\lVert \hat{\xi_j} \right\rVert_1,
\label{eq:5}
\end{equation}
where $\dot{W}_j$ indicates the time derivative of the $j^{th}$ row of $\dot{\mathbf{W}}$. Similarly, $\xi_j$ denotes the $j^{th}$ row of $\Xi$. The parameter $\lambda$ acts as a sparsity-promoting factor, aiding in the selection of a sparse $\Xi$ with the fewest library terms, thereby enhancing the model's efficacy.

 The determination of the optimal number of clusters is guided by the F-test criterion, which simultaneously influences the setting of the embedding dimension. Notably, our proposed cluster regression model inherently determines both the embedding dimension and the time lag between delay coordinates, anchored by the chosen cluster resolution $K$.

\subsection{Model predictive control}

To harness the cluster regression model for control, it is integrated with the Model Predictive Control (MPC) methodology \cite{allgower1999nonlinear,camacho2007model}. MPC offers a dynamic optimization technique that incorporates input constraints into the control strategy. In the context of the cluster regression model, every state is termed a cluster state. Consequently, the terms "current state" and "target state" in control parlance correspond to the "current cluster state" and "target cluster state," respectively.

The MPC operates by predicting future states using the cluster regression model for a span of \(n_p\) steps, known as the prediction horizon. Subsequently, an optimal sequence of control inputs is determined for \(n_c\) steps, termed the control horizon. Typically, the control horizon does not exceed the prediction horizon. The formulation of the optimal control inputs, \(\mathbf{u}(\mathbf{S}_j) = [\mathbf{u}_{j+1},\cdots,\mathbf{u}_{j+n_c}]\), for the current state estimate \(\mathbf{S}_j\) at time step $j$ is anchored in the minimization of a cost function, \(J\):
\begin{equation}
J = \sum_{k=0}^{n_p-1} ||\mathbf{S}_{j+k} - \mathbf{T}_{k}||_\mathbf{Q}^2 + \sum_{k=1}^{n_c-1} \left(||\mathbf{u}_{j+k}||_{\mathbf{R}_u}^2 + ||\Delta \mathbf{u}_{j+k}||_{\mathbf{R}_{\Delta u}}^2\right),
\end{equation}
where \(\mathbf{T}_k\) denotes the target state at step $k$ and \(\Delta \mathbf{u}_k = \mathbf{u}_k - \mathbf{u}_{k-1}\) signifies the rate of change in the control input. In this formulation, the weight matrix \(\mathbf{Q}\) imparts penalties for deviations from the reference trajectory. Simultaneously, \(\mathbf{R}_{u}\) and \(\mathbf{R}_{\Delta u}\) penalize the control cost and rate of input change, respectively.

\section{Results}
\label{results}

\subsection{Lorenz system}

The Lorenz system \cite{lorenz1963deterministic} is a set of differential equations originally derived from a model of atmospheric convection. These equations describe chaotic dynamics and have been the subject of numerous studies since their introduction. When an external forcing term is introduced, the system can be described by:
\begin{subequations}
\begin{align}
\frac{dx}{dt} &= \sigma (y - x) + u \\
\frac{dy}{dt} &= x(\rho - z) - y \\
\frac{dz}{dt} &= xy - \beta z,
\end{align}
\end{subequations}
where the system is forced exclusively in the $x$ direction. The parameter values are set as $\sigma = 10$, $\beta = \frac{8}{3}$, and $\rho = 28$ for the analysis below. 

Utilizing time series data for the variables \( x \), \( y \), and \( z \) (for detailed information, refer to Appendix \ref{appendixD1}), we construct a three-dimensional feature space to visualize the system's dynamic behavior. The Lorenz system inherently exhibits oscillations between two weakly stable fixed points, approximately located at \( (\pm \sqrt{72}, \pm \sqrt{72}, 27)^T \). This oscillatory pattern, while predictable over short intervals, becomes increasingly unpredictable over extended periods due to the system's sensitivity to initial condition - a characteristic feature of chaotic systems. Such sensitivity introduces substantial challenges in both the identification and control of the system \cite{kaiser2018sparse}, necessitating advanced modeling and control strategies to effectively understand and manage its complex dynamics.

\begin{figure}[t!]
    \centering
    \includegraphics[width=0.9\textwidth]{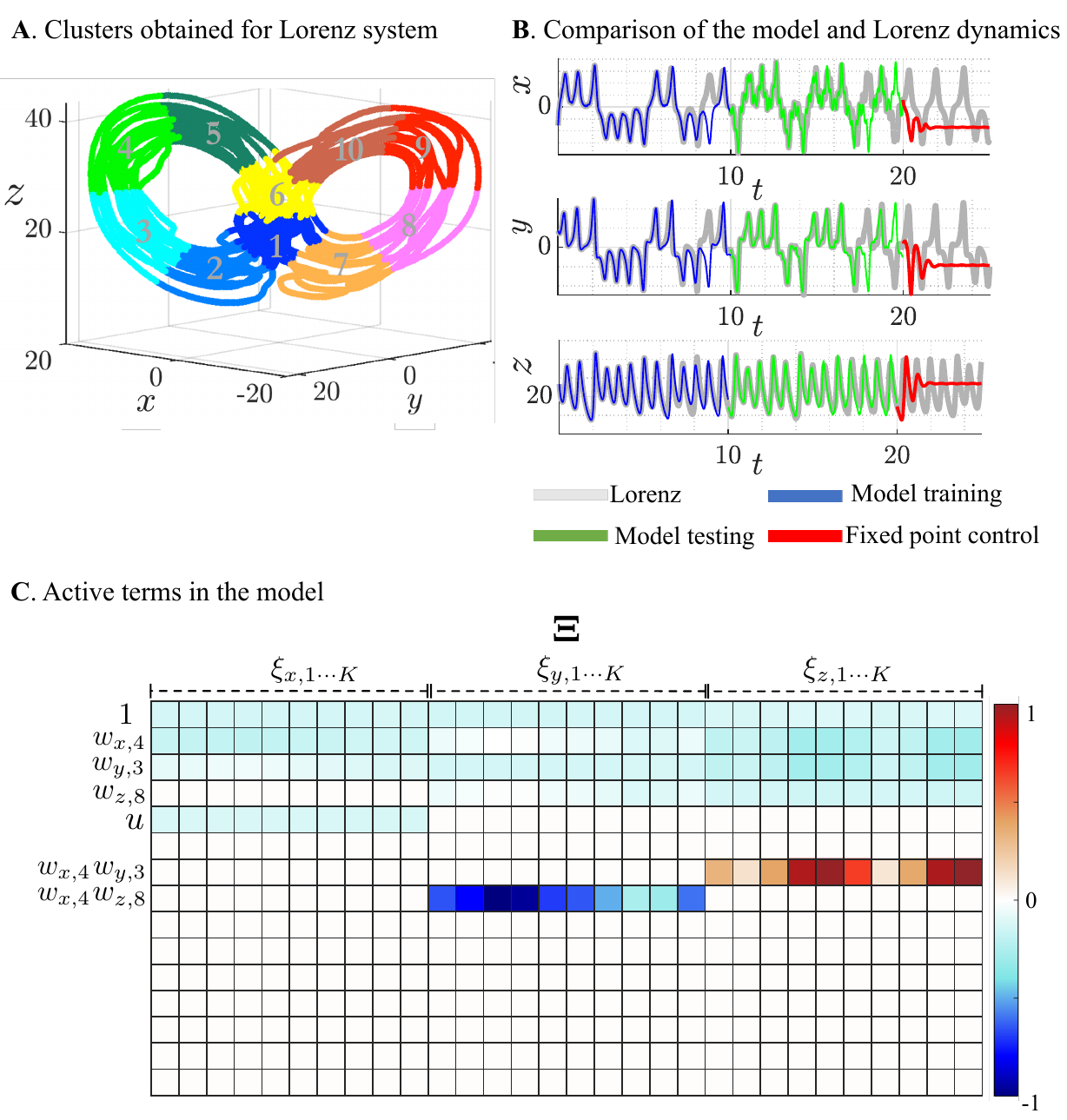}
    \caption{Model construction for the forced Lorenz system. \textbf{A}. Clustering analysis: Dissecting the butterfly-shaped attractor into $K=10$ distinct clusters. \textbf{B}. Model prediction: Juxtaposition of model predictions with the forced Lorenz dynamics during both training ($t=0$ to $t=10$) and testing phases ($t=10$ to $t=20$), along with a demonstration of equilibrium point control. \textbf{C}. Model analysis: Identification of 15 linearly independent columns from the library function $\Theta$. Emphasis on the most dominant terms in the model, highlighting dependencies on both linear and quadratic terms. Terms are scaled between -1 and 1 to facilitate interpretation.}
    \label{fig:2}
\end{figure}

Fig. \ref{fig:2}(A) illustrates a clustering of the Lorenz attractor into $K=10$ clusters, determined using an F-test criterion (see Appendix \ref{appendixD2} for details). On closer inspection, these clusters can be grouped into three main categories: the clusters $2$, $3$, $4$, and $5$ which correspond to the left lobe of the attractor; the clusters $7$, $8$, $9$, and $10$ which represent the right lobe; and clusters $1$ and $6$ which function as transitional clusters facilitating movement between the two lobes \cite{kaiser2014cluster}. These transitional clusters play a crucial role, especially when a trajectory initiated from a particular state tends to remain within one lobe for a certain duration (see Appendix \ref{appendixD3} and  \ref{appendixD4} for details). However, due to the inherent chaotic nature of the system, it eventually transitions to the other lobe via these transitional clusters. One complementary approach to predict such lobe switching involves representing the Lorenz system's chaotic behavior as a forced linear system \cite{brunton2017chaos}.

For predictive modeling, the Lorenz system utilizes polynomial library functions up to an order of $p=2$. As previously discussed, the QR decomposition with column pivoting identifies 15 dominant columns. Fig. \ref{fig:2}(C) displays the active coefficients for this model. The $x$ features are associated only with linear terms, while the $y$ and $z$ features include both linear and quadratic terms in their coefficients.

The predictive model undergoes training for the initial 10 seconds, followed by a validation phase lasting another 10 seconds. Over the subsequent 5 seconds, MPC is applied to control the trajectory (refer to Appendix \ref{appendixD5} for a more details). The overarching control objective aims to stabilize the trajectory around one of the system's fixed points. Throughout both the training and validation stages, the model adeptly mirrors the dynamics of the forced Lorenz system, as demonstrated in Fig. \ref{fig:2}(B). It's also evident that the controller efficiently steers the trajectory towards a fixed point, showcasing the robustness of the model (comparative studies are detailed in Appendix \ref{appendixD6}). The sensitivity of the cluster regression model to training time as well as control towards to a fixed point with different initial conditions is discussed in Appendix \ref{appendixD7}.

\subsubsection*{Control of Lorenz system}

Leveraging the control architecture previously discussed, we further delve into the control aspects of the Lorenz system. The equilibrium point control, when integrated with the current model, demonstrates remarkable efficiency. Here, we elucidate some strategies to effectively navigate the system through various cluster states.

\begin{figure}[t!]
    \centering
    \includegraphics[width=0.9\textwidth]{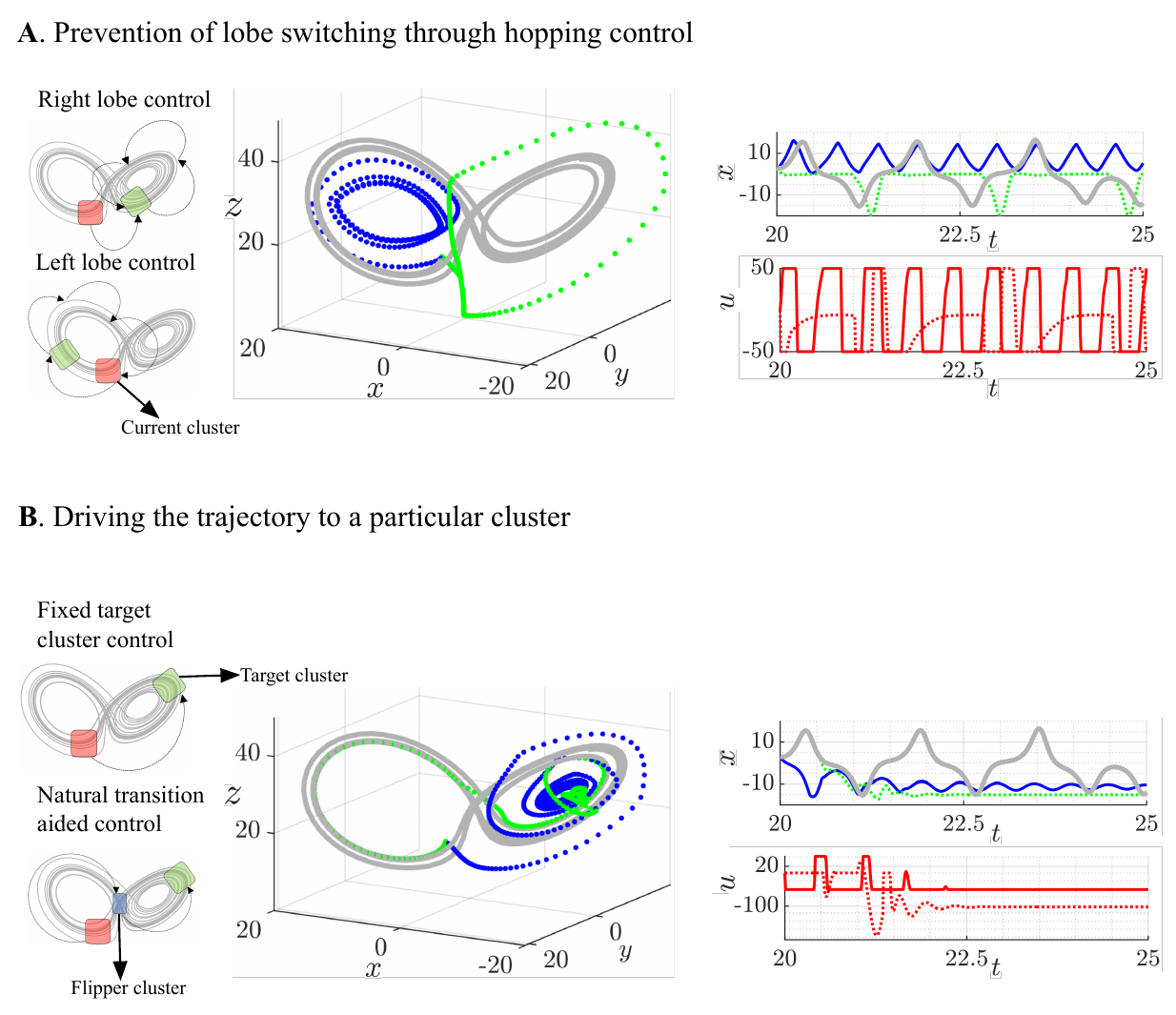}
    \caption{Controlling the Lorenz system dynamics. \textbf{A}. Cluster hopping concept: Illustration of trajectory guidance within the Lorenz system. Trajectories are either directed and sustained within the right lobe (green) or confined to the left lobe (blue), demonstrating controlled movement between clusters. \textbf{B}. Trajectory steering: Employing targeted control tactics (blue) and leveraging the intrinsic dynamics of the system to navigate the trajectory towards a specified cluster (green), showcasing effective manipulation of the system's path. The dotted red lines for control input $u$ correspond to the green trajectories and solid red lines for $u$ correspond to the blue trajectories in both \textbf{A} and \textbf{B}. The grey lines indicate the baseline Lorenz dynamics.}
    \label{fig:3}
\end{figure}

Fig. \ref{fig:3}(A) captures the concept of `cluster hopping', where the trajectory is systematically driven from one cluster to the next by iteratively updating the source (shown in the red box) and target clusters (shown in the green box) within the MPC framework. This `hop-and-switch' approach highlights the controller's ability to fluidly transition between cluster states.

By employing cluster hopping, we demonstrate two scenarios: in the first, the trajectory remains predominantly within the right lobe (denoted in green, labeled as right lobe control), while in the second, it stays confined to the left lobe (indicated in blue, labeled as left lobe control). In both scenarios, a distinct frequency shift in the system's dynamics is evident. This underscores the controller's capability to modulate system frequencies while adhering to intrinsic dynamics, a feature of paramount importance in controlling diverse multiphysics systems. For instance, in applications such as mitigating aeroelastic or aeroacoustic resonances, it becomes crucial to selectively dampen certain frequencies that could instigate a feedback loop.

We also examine the trajectory derived from guiding the system between a predefined source and a target cluster. As illustrated in Fig. \ref{fig:3}(B), our intent is to navigate the trajectory from its source (cluster $2$) directly to the target (cluster $9$). Considering the distinct positioning of the target cluster on a different lobe of the attractor, our controller proactively diverts the trajectory (highlighted in blue) away from the initial attractor to facilitate the transition. Notably, reaching the target cluster isn't immediate. The system's inherent dynamics initially guide the trajectory around the right lobe before converging, in a manner reminiscent of fixed-point control, spirally toward cluster $9$.

Efficient control design often mandates a nuanced understanding of the system's intrinsic dynamics. Ideally, control should be applied at pivotal junctures to optimize trajectory transitions. For example, in the context of the current system, to transition from cluster $2$ to cluster $9$, it becomes evident that cluster $1$ acts as a natural intermediary step called as the flipper cluster. This is because, without any external forcing, the system typically progresses through a sequence of clusters: $2 \rightarrow 3 \rightarrow 4 \rightarrow 5 \rightarrow 6 \rightarrow 1$, a progression attributed to its intrinsic dynamics (detailed analysis can be found in Appendix \ref{appendixD3}). After reaching the flipper cluster, the controller can then be employed to guide the trajectory to cluster $9$. This strategic approach is depicted in Fig. \ref{fig:3}(B) with the trajectory shown in green. It's noteworthy that up until the trajectory reaches the flipper cluster, its behavior mimics the unforced system. Upon activation of the control, there's a swift transition to the target cluster. This methodology underscores the importance of understanding state transitions within the system, which can be instrumental in devising control strategies tailored to specific system regimes.

\begin{figure}
    \centering
    \includegraphics[width=0.85\textwidth]{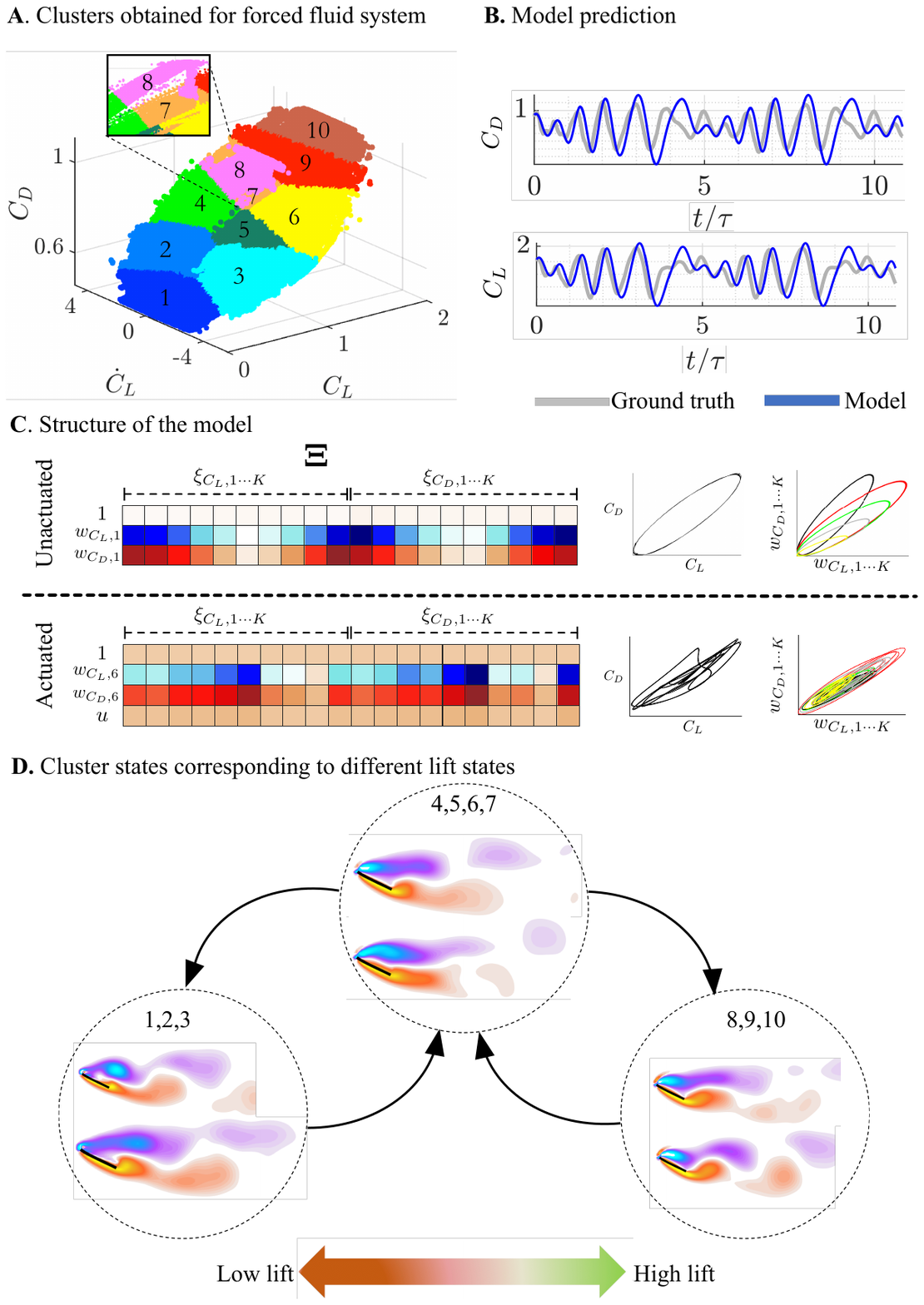}
    \caption{Model dynamics of flow over a flexible flat plate at an angle of attack of $35^\circ$. \textbf{A}. Cluster representation: Depicting the trajectories of flow over the flexible flat plate, categorizing flow patterns into distinct clusters for detailed analysis. \textbf{B}. Model prediction: Conducting a comparative analysis between the cluster regression model predictions and the forced dynamics obtained through direct numerical simulation (DNS), demonstrating a high degree of accuracy. \textbf{C}. Model structure: (left) Identification of active model terms (values are normalized between $-1$ and $1$ for visualization purposes). (right) portraits of $C_L$, $C_D$ and corresponding cluster coefficients (colors denote different pairwise cluster coefficients). \textbf{D}. Feature clustering delineates distinct lift (and drag) states: The vorticity contours corresponding to three distinct lift states (determined through community detection) are showcased, with arrows indicating transitional pathways among them.}
    \label{fig:4}
\end{figure}

\subsection{Flow over a flexible flat plate}

We now turn our attention to a canonical fluid-structure interaction system in a high-dimensional setting. Through direct numerical simulation (DNS), we study an incompressible flow over a flexible flat plate positioned at a $35^\circ$ angle of attack (see Appendix \ref{appendixE1} for details of the numerical simulation). At a low Reynolds number of $Re=100$, the incident flow induces a regular vortex shedding pattern, caused by flow separations at both leading and trailing edges. This results in periodic oscillations of lift and drag forces on the plate. Notably, the fluctuations in lift and drag are synchronized. 

Momentum is injected close to the plate's leading edge normal to the flow, providing an unsteady external input to the flow. This jet velocity, utilized for actuation, serves as the cluster regression model's forcing term (see Appendix \ref{appendixE1} for more details). The introduction of this force modifies the periodic vortex shedding-induced limit cycle oscillations in two significant ways: by directly modulating the flow circulation behind the plate, either amplifying or diminishing it; and by causing movement in the plate, which can pivot around its hinged leading edge and flex due to its inherent pliability.

In our cluster regression approach to model this system, we rely on time-series data of the lift coefficient, its time derivative, and the drag coefficient as key features \cite{nair2019cluster,taira2018phase, loiseau2018sparse}. The lift coefficient and its derivative adeptly depict the amplitude and phase dynamics of the oscillations, while the drag coefficient highlights the time-averaged flow's deviation from the unstable steady state.

We partition the feature space into $K = 10$ clusters, as illustrated in Fig. \ref{fig:4}(A) (for more specifics, refer to Appendix \ref{appendixE2}). The clusters densely span the range from the maximum to the minimum lift and drag states. It's worth highlighting that, in the case of unforced limit cycle oscillations, the clustering process differentiates the feature space based on phase. Hence, the dense coverage in this context indicates a rich, predominantly non-periodic feature space suitable for model training. We devise a cluster regression model targeting the lift and drag features. It's noteworthy that constructing a model for $C_L$ intrinsically encompasses a model for $\dot{C}_L(t)$. The alignment between our model's prediction and the true DNS data measurements is depicted in Fig. \ref{fig:4}(B).

For the system without external forcing input, an examination of the top panel in Fig. \ref{fig:4}(C) reveals that the model seizes a single coefficient for each feature space, displaying independence from the constant term. Delving deeper, there's a discernible symmetry in the dynamic coefficients related to $C_D$ and $C_L$. To elucidate, coefficients $\xi_{C_D,1}$ and $\xi_{C_L,1}$ provide analogous weightage for the prediction of cluster coefficients $w_{C_L,1}$ and $w_{C_D,1}$ respectively. Conversely, when the system is influenced by an external force, the dynamic coefficients for the model are showcased in the lower segment of Fig. \ref{fig:4}(C). What stands out here is our attainment of a sparse predictive model dominated by active linear terms. Taking the flow's unstable steady-state progression into consideration, one might intuit the emergence of active cubic nonlinearities in the model \cite{loiseau2018sparse, noack2003hierarchy}. Yet, diverging from the unforced scenario, the coefficients lack symmetry. Additionally, there's a noticeable presence of an active constant and forcing term in the model. The evolution narrative for each cluster coefficient is threaded by a constant term $1$, the forcing component $u$, one cluster coefficient attributed to lift $w_{C_L,6}$, and one to drag $w_{C_D,6}$.

The right segment of Fig. \ref{fig:4}(C) functions as a platform to display the phase portraits of the forecasted cluster coefficients for both unactuated and actuated systems. The trajectories of the pairwise cluster coefficients, such as \( (w_{C_L,1}, w_{C_D,1}) \), \( (w_{C_L,2}, w_{C_D,2}) \), and others, reflect the dynamic nature observed in the phase portrait of the primary lift and drag coefficients, which are visualized alongside. Various colors represent different pairwise cluster coefficients, highlighting the diversity in the system's response. This parallelism underscores that, regardless of the cluster coefficients, a consistent structural pattern should be evident in the model. For example, in the unactuated case, a symmetry is observed where each pair of coefficients converges to a limit cycle, as illustrated in the right plot. This observation confirms that a single cluster coefficient for each dimension of the feature space can effectively encapsulate the temporal dynamics of the system.

The low-dimensional cluster representation offers a clear visualization of the high-dimensional system's state. For example, to discern flow structures responsible for low drag, one can examine snapshots aligned with low drag states. With a predictive model in place for cluster states, the cluster regression model articulates the high-dimensional, non-linear system through its evolution equation. Such models are optimal for analyzing and controlling a broad spectrum of dynamical systems.

Employing a community detection algorithm \cite{newman2006modularity}, clusters are categorized into three groups: low-lift, mid-lift, and high-lift states. Any given flow state over time can be viewed as a transition between these clusters. For instance, Fig. \ref{fig:4}(D) showcases a transition from low-lift clusters through mid-lift clusters, culminating in high-lift clusters. While transition diagrams might vary across systems, they underscore an essential concept: some states could become inaccessible if intermediary cluster states are bypassed. This insight can lead to innovative control designs tailored to specific applications. Fig. \ref{fig:4}(D) exhibits the high-dimensional snapshots (represented by vorticity contours) corresponding to these distinct states, underscoring the flow structures that shape these cluster states. The capability to vividly visualize these states is immensely advantageous, facilitating nuanced analysis and the design of efficient closed-loop controls discussed next. More details on the transition matrix and cluster importance for this high-dimensional system is provided in Appendix \ref{appendixE3} and  \ref{appendixE4}.

\begin{figure}
    \centering
    \includegraphics[width=0.9\textwidth]
    {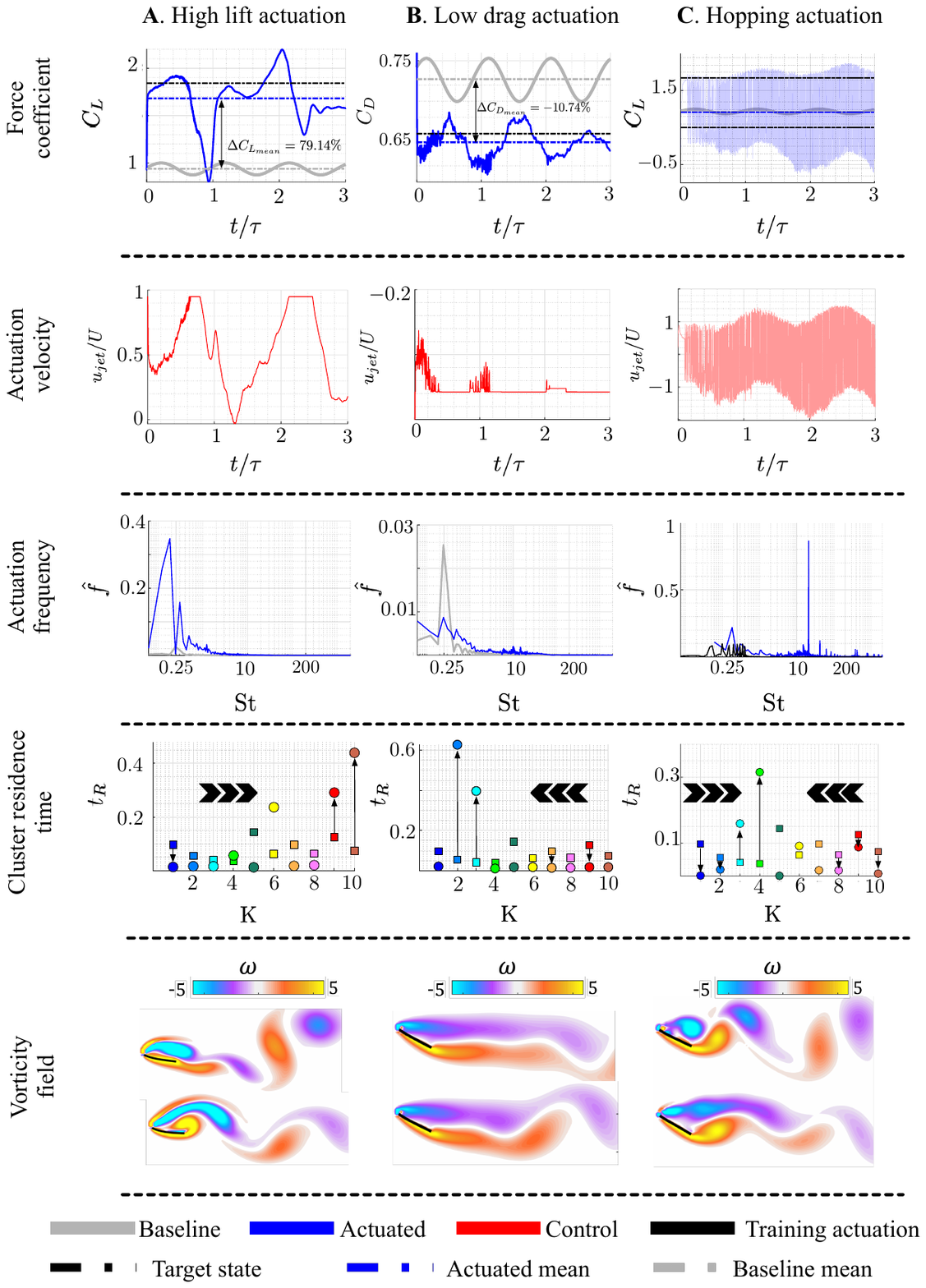}
    \caption{Flow control characteristics with cluster regression model for \textbf{A}. High lift actuation, \textbf{B}. Low drag actuation and $\mathbf{C}$. Hopping actuation. For each actuation strategy, we highlight (top row) the aerodynamic force coefficients with actuation, (second row) the forcing input required for actuation, (third row) the actuation frequency compared with baseline shedding frequency and training actuation (in grey), (fourth row) the cluster residence time for actuation (circles) compared with baseline cases (squares), and (fifth row) the vorticity contours with actuation at two different times.}
    \label{fig:5}
\end{figure}

\subsubsection*{Flow control}

In our study, we primarily aimed to modify the circulation across the plate using an actuator at the leading edge, which either injects or extracts momentum normal to the incoming flow. Generating high lift through jet injection is particularly suitable for bio-inspired micro air vehicles, given their structural constraints against flapping. In gust-prone environments, it's crucial to manage adverse lift transients by tempering the circulation over the wing. We explore both these methodologies in our work using a model-predictive control setup (see Appendix \ref{appendixE5} for more details). Furthermore, we evaluate the controller's aptitude for rapid transitions between desired clusters, termed as cluster hopping. This is vital in gusty conditions due to the brief timescales of gust transients \cite{jones2022physics}. For every scenario, a baseline unactuated case undergoes control for three convective cycles.

When steering the flow state towards high lift cluster state $10$, there's a $79\%$ surge in the mean lift coefficient as depicted in the first row panel of Fig. \ref{fig:5}(A), with a momentum coefficient $C_\mu = 0.0057$ (defined in Appendix \ref{appendixE1}). Positive actuation values, as observed in the second panel of Fig. \ref{fig:5}(A), indicate momentum injection enhancing the lift. This increased lift, as evidenced by the fourth row panel of Fig. \ref{fig:5}(A), lengthens the residence time in high lift clusters, suggesting the flow predominantly assumes high lift states . Vorticity contours, presented in the fifth row panel of Fig. \ref{fig:5}(A), further emphasize dominant vortices atop the plate during these conditions.

Aiming for a low drag state of cluster $2$ ensures the average drag consistently remains below the unaltered baseline, as illustrated in the first row panel of Fig. \ref{fig:5}(B). This actuation culminates in a $10.74\%$ drop in mean drag with a momentum coefficient $\bar{C}_{\mu} = 0.0019$. The actuation profile for this scenario, highlighted in the second row panel of Fig. \ref{fig:5}(B), displays momentum extraction, decreasing plate circulation throughout. With cluster $2$ as the intended low drag cluster, its residence time, alongside cluster $3$, augments due to the flow's persistent low drag inclination (see Appendix \ref{appendixE6} for dominant transitions due to the actuation). This is mirrored in the vorticity contours of the fifth row panel of Fig. \ref{fig:5}(B), showcasing extended leading-edge vortices symbolic of low drag states.

For scenarios demanding swift state changes and immediate stabilization, controllers must adeptly alternate between states. The hopping control fulfills this, dynamically adjusting source and target clusters and ensuring rapid transitions between them. This is evident as the lift coefficient, seen in the first row panel of Fig. \ref{fig:5}(C), oscillates between peak and nadir target clusters throughout the sequence. The actuation velocities, detailed in the second row panel of Fig. \ref{fig:5}(C), encompass both momentum injection and extraction, influencing plate circulation accordingly. Interestingly, the actuation frequency in this mode, as shown in the third row panel of Fig. \ref{fig:5}(C), surpasses the model's training frequency. For the other two scenarios, as seen in the third row panels of Fig. \ref{fig:5}(A) and \ref{fig:5}(B), it aligns with the training actuation and the shedding frequency of the baseline. 

In Fig. \ref{fig:5}(C), the flow gravitates more towards medium lift states since any movement to high lift or low drag states prompts the controller to counteract promptly. This intensifies the residence time in medium lift clusters. Owing to the swift transitions between high lift and low drag states, the vorticity field in this mode comprises diverse flow structures emerging over concise intervals. While the hopping control doesn't majorly deviate from the mean lift coefficient, it does heighten its variance. Significantly, even as actuation modulates vortex shedding frequency, the average lift force on the plate remains stable which is in stark contrast to methods such as phase-based control \cite{nair2021phase} where exercising control over the vortex shedding frequency induces alterations in force coefficients. This novel control approach holds promise in fields like energy harvesting, vortex-induced vibration-based heat transfer enhancements, and medical innovations like micro-fluidic vortex shedding for controlled gene delivery \cite{wang2020state, lee2019vortex, izadpanah2019effect, jarrell2019intracellular}. The cluster regression model and control performance at higher Reynolds number is further discussed in Appendix \ref{appendixE7}.

\section{Discussion}
\label{conclusions}

We have introduced a straightforward, data-driven model that offers a compressed representation and control of high-dimensional systems using minimal sensor measurements. This makes it apt for data-derived from both experimental and computational sources. Our framework provides an efficient and easily implementable solution for controlling intricate systems. At its core, the model utilizes data clustering, which segments vital system information into distinct clusters. These clusters can, in turn, inform innovative control strategies. The model has showcased exceptional efficacy in managing high-dimensional systems.

However, the model is not without its limitations, primarily stemming from the clustering process. One major concern is determining the number of cluster centroids. Although the optimal cluster count can be ascertained using the elbow criterion (refer to Appendix \ref{appendixB1}), our model's representation hinges on achieving an adequate system portrayal. This entails assessing if adjusting the cluster count significantly alters cluster states. Another challenge is the dependency of cluster centroids on their initial values. We can mitigate this by employing kmeans++ \cite{arthur2007k}, as done in our work, which judiciously selects initial centroids. But even with this method, the clustering results can vary with different iterations on certain datasets, potentially leading to inaccurate model outcomes.

It is crucial to note that external forces, which modify system dynamics, can influence the clustering. For simpler systems like the Lorenz system, forced dynamics generally revolve around the butterfly manifold, making changes in clustering less pronounced. But for more complex systems, like turbulent flows, there's a marked difference between unforced and forced dynamics. With substantial external forces, the model might deviate significantly from the training data, leading to implausible predictions. We aim to address this in future work by enabling the model to learn and adjust to cluster changes during control using active learning \cite{mania2022active}. In our current study, the model operated at frequencies surpassing those of training actuation. We modulated the forcing in part to maintain reliable model predictions and prevent the onset of instabilities in our numerical solutions. Looking ahead, we aim to apply this methodology to more demanding scenarios, such as turbulent flows around flexible structures experiencing nonlinear flutter oscillations.

\section*{Acknowledgements} 

AGN acknowledges the support from the Department of Energy Early Career Research Award (Award no: DE-SC0022945, PM: Dr. William Spotz), the National Science Foundation AI Institute in Dynamic systems  (Award no: 2112085, PM: Dr. Shahab Shojaei-Zadeh) and US DoD/US Department of the Air Force (Award number FA9550-23-1-0483, PM: Gregg Abate).

\begin{appendix}

	\setcounter{figure}{0}
	\setcounter{equation}{0}
	\renewcommand\thefigure{S\arabic{figure}} 
	\renewcommand\theequation{S\arabic{equation}}

\section{Technical background}
\label{appendixA}

In the pursuit of advancing airborne vehicles, ranging from diminutive drones to expansive civilian aircraft, there has been a concerted push towards achieving greater efficiency and reduced noise levels. This drive has fostered considerable investigation into active flow control methodologies \cite{cattafesta2011actuators,greenblatt2022flow}. While some studies advocate the use of open-loop forcing as a mechanism to modulate dominant frequencies or leverage non-linear frequency interactions \cite{greenblatt2000control}, there exists a distinct preference towards closed-loop control within the community. This preference is anchored in the notion of predicting system dynamics to derive an effective control strategy \cite{brunton2015closed}.

Confronted with the intrinsic high-dimensionality of fluidic systems, the scientific community recognizes the necessity of distilling these systems into more manageable, low-dimensional representations, thus setting the stage for the formulation of reduced-order models \cite{noack2011reduced}. A venerated technique to this end involves the transformation of high-dimensional data snapshots into coherent, low-dimensional modes through proper orthogonal decomposition (POD) \cite{berkooz1993proper,holmes2012turbulence}. The temporal dynamics of these modes, as encapsulated by the POD coefficients, can be elucidated by projecting the Navier--Stokes equations onto these modes, culminating in a Galerkin or Petrov-Galerkin system \cite{antoulas2005approximation,rowley2017model}. Among the cadre of POD-inspired reduced-order models (ROMs) tailored for linear system control, the balanced POD stands out, underpinned by the calculation of empirical Gramians to effectuate an approximate balanced truncation \cite{rowley2005model, moore1981principal}.

However, the architectural design of ROMs is not solely confined to projection methods. Modern approaches exhibit the feasibility of deriving ROMs exclusively from data. Classic models inspired by the works of Wagner \cite{wagner1924entstehung} and Theodorsen \cite{theodorsen1949general}, which incorporate the nuances of viscous forces, exemplify this trajectory. These models have been dexterously deployed to control aerodynamic forces, particularly in scenarios typified by low Reynolds number flows \cite{brunton2013empirical,brunton2013reduced,brunton2014state}. Extending this paradigm, recent innovations account for wing flexibility, crucial for dictating aggressive maneuvers \cite{hickner2023data}. In parallel, researchers have explored autoregressive models like ARMAX \cite{herve2012physics} and NARMAX \cite{billings2013nonlinear,fabbiane2014adaptive}, the eigensystem realization algorithm which relies on impulse responses for system identification \cite{juang1985eigensystem}, dynamic mode decomposition (DMD) that yields pure frequency modes \cite{rowley2009spectral,schmid2010dynamic}, and network-theoretic models aimed at deciphering vortex interactions \cite{nair2015network}. Kaiser et al. \cite{kaiser2014cluster} introduced an intriguing cluster-based reduced order model, a blueprint we draw inspiration from in our current endeavor.

Reflecting on clustering methodologies, one of the pioneering works utilized the Centroidal Voronoi Tessellation (CVT) to generate a reduced order model for the Navier-Stokes equation \cite{burkardt2006pod}. Subsequent studies, including those by Kaiser et al. \cite{kaiser2014cluster}, evolved this concept by adopting the k-means clustering technique and leveraging a probabilistic Markov model to depict transitions between clusters. The cluster-reduced order model (CROM) derived from this methodology has been instrumental in a range of applications. It facilitated the extraction of pertinent features in the context of a mixing layer. This model further proved invaluable in probing cycle-to-cycle fluctuations in internal combustion engines \cite{cao2014cluster}, enabling unsupervised categorization of actuated flows \cite{ishar2019metric}, and elucidating the transition dynamics inherent in supersonic mixing layers \cite{li2020cluster}. Furthermore, integrating control designs within this framework, feedback control strategies have been developed to diminish the mean recirculation area over smooth ramps \cite{kaiser2016cluster}. Additionally, a model-independent approach has been devised to modify cluster transitions, which has been particularly effective for drag reduction in post-stall airfoil configurations \cite{nair2019cluster}. 

While the cluster-reduced order model (CROM) has proven advantageous in several scenarios, the dynamics rely on the progression of an ensemble of trajectories denoted by a state probability vector. However, this state probability tends to rapidly converge to its asymptotic state, making CROM less viable for extended predictions or certain model-based control designs. Li et al. \cite{li2021cluster} endeavored to enhance the dynamic component of CROM, introducing a cluster-based network model (CNM). In this model, dynamics are constrained to a sparse network connecting cluster centroids. A key premise was that the transition time between centroids would reflect the mean residence time of both clusters involved. Yet, the assumption of consistent transition times in CNM introduces a somewhat arbitrary constraint. 

Our current approach leverages system identification techniques \cite{brunton2016discovering,brunton2016sparse} to achieve a streamlined representation of system dynamics that encompasses control inputs. This framework can be effortlessly adapted to account for physical constraints \cite{loiseau2018constrained}. Oftentimes, direct access to high-dimensional data is limited, necessitating models built on limited sensor readings \cite{kaiser2018sparse}. Such data either provides a comprehensive state estimate \cite{loiseau2018sparse} or helps discern inherent dynamics \cite{floryan2022data}. In this study, we present a cluster regression model leveraging limited sensor data—specifically, lift and drag coefficients from direct numeric simulations (DNS) examining laminar incompressible flow over a compliant flat plate at a $35^{\circ}$ attack angle. While the setup draws inspiration from insect flight \cite{mountcastle2013wing, akkala2015vortex,eldredge2019leading}, it aptly illustrates fluid-structure interactions observed in flexible wind turbine blades \cite{hsu2012fluid}, flutter mitigation in gusty conditions \cite{menon2020aeroelastic}, and energy harvesters \cite{mathai2022fluid}.

\section{Methods}
\label{appendixB}

\subsection{Data-driven Clustering}
\label{appendixB1}

Data classification into distinct groups is pivotal in many scientific and engineering domains, aiding in pattern recognition and data compression. Clustering, particularly centroid-based clustering, is a key analytical tool for identifying subsets within a dataset based on similarity in observations, often assessed using distance metrics.

Centroid-based clustering, exemplified by the k-means algorithm \cite{lloyd1982least}, partitions a dataset into \( K \) clusters, \( \mathcal{C}_k, k=1, \ldots, K \), each linked to a unique centroid. Initially, measurements \( x_1(t), x_2(t), \ldots, x_n(t) \) are gathered over time. These are arranged into a vector \( \mathbf{q} = [x_1(t) \; x_2(t) \ldots x_n(t)] \) to form an n-dimensional feature space. The centroid of a cluster \( \mathcal{C}_k \), \( \mathbf{c}_k = [c_{1,k} \; c_{2,k} \ldots c_{n,k} ] \), is computed as the mean of all observations in that cluster:

\begin{equation}
	\mathbf{c}_k = \frac{1}{N_k} \sum_{\mathbf{q}_m \in \mathcal{C}_k} \mathbf{q}_m.
\end{equation}
Here, \( N_k \) is the number of points in cluster \( k \), and \( \mathbf{q}_m \) denotes the \( m^{th} \) observation. This unweighted mean method aligns with centroidal Voronoi Tessellation approaches \cite{du1999centroidal,burkardt2006centroidal}.

In k-means clustering, initial centroid selection is crucial as it influences the final outcome. This issue can be mitigated through multiple initializations or intelligent seeding methods like kmeans++ \cite{arthur2007k}. The algorithm allocates each data point to the nearest cluster, then recalculates centroids based on these assignments. This iterative process of reallocation and centroid recalculation continues until changes fall below a predefined threshold. The goal is to minimize the intra-cluster variance \( J_w \):

\begin{equation}
	J_w = \frac{1}{N}\sum_{k=1}^{K} \sum_{\mathbf{q}_m \in \mathcal{C}_k} \|\mathbf{q}_m - \mathbf{c}_k\|^2.
\end{equation}

This process effectively reduces the high-dimensional data space into a set of centroidal Voronoi tessellations. Each centroid encapsulates the averaged attributes of its cluster. In periodic systems, this means that the centroid is the phase average of the encompassing measurements, discretizing the state space into distinct cluster characteristics. However, as Burkardt et al. \cite{burkardt2006centroidal} showed, selecting the optimal number of clusters is key to maintaining representational fidelity without incurring significant cluster variance loss.

The elbow method \cite{hartigan1975clustering,tibshirani2001estimating} helps identify the ideal number of clusters, marked by the inflection point on a metric-vs.-clusters plot. Common metrics include Bayesian Information Criterion (BIC) \cite{schwarz1978estimating} and the F-test \cite{lomax2013introduction}. The BIC in k-means is expressed as:

\begin{equation}
	BIC = \sum_{k=1}^{K} \left\{- \frac{1}{2}N_k \log|\Sigma_k|\right\} - M K \left(n + \frac{1}{2}n(n+1)\right),
\end{equation}

where \( \Sigma_k \) is the sample covariance matrix of cluster \( k \), \( M \) the total number of measurements, and \( n \) the dimensionality of the feature space. The F-test calculates the inter-cluster variance \( J_i \) as:

\begin{equation}
	J_i = \frac{1}{N} \sum_{k=1}^{K} N_k \|\mathbf{c}_k - \bar{\mathbf{c}}\|^2,
\end{equation}

with \( N_k \) being the count of measurements in cluster \( \mathcal{C}_k \), and \( \bar{\mathbf{c}} \) the overall centroid. The chosen metric, the ratio \( J_i/J_T \) of inter-cluster to total cluster variance, indicates an optimal cluster count when an elbow appears. In this study, an elbow at \( J_i/J_T = 0.9 \) suggests choosing \( K=10 \) clusters, used in all simulations unless otherwise noted.

\subsection{Limitation of previous cluster-based reduced-order models}
\label{appendixB2}

The discretization of state space into \( K \) optimal clusters enables the description of dynamics within these clusters, akin to a Proper Orthogonal Decomposition (POD) based reduced order model \cite{noack2011reduced}. Unlike POD models, which focus on the dynamics of temporal coefficients, cluster-based reduced order models (CROMs) elucidate transitions among clusters. At each time step, trajectories originating within the same cluster may diverge. Kaiser et al. \cite{kaiser2014cluster} developed a dynamic model using a state probability vector \( \mathbf{p} = [p_1, p_2, \ldots, p_K]^T \), where \( p_k \) signifies the probability of the system being in cluster \( k \). Transitions between clusters are represented by a transition probability matrix \( \mathbf{P} \in \mathbb{R}^{K \times K} \), with \( P_{ij} \) indicating the transition probability from cluster \( j \) to \( i \), propagating the state probability vector as:

\begin{equation}
	\mathbf{p}^{t+1} = \mathbf{P}\mathbf{p}^t.
\end{equation}

CROMs assume Markovian behavior, suggesting future states depend only on the current state, similar to the Perron-Frobenius operator's evolution for an ensemble of trajectories. However, this reliance on the Markov property can cause a diffusion of the state probability, reducing predictive accuracy over time. Li et al. \cite{li2021cluster} addressed this by introducing a cluster-based network model, eliminating time-step dependencies and hypothesizing average transition times between clusters.

The expectation value for the \( i^{th} \) feature space variable in CROMs is expressed as:
\begin{equation}\label{eq:6}
	x_i = \sum_{k=1}^{K} p_{k}c_{i,k} \quad \text{for } i = 1, 2, \ldots, n,
\end{equation}
where \( c_{i,k} \) is the centroid of the \( k^{th} \) cluster for the \( i^{th} \) feature space dimension. However, once the state probability vector stabilizes, this expected value may not accurately reflect the actual system dynamics. In contrast, the cluster network model, for \( t \) within \( [t_n, t_{n+1}] \), defines the feature space variable as:

\begin{equation}
	x_i = \alpha_n(t)c_{i,k_n} + [1 - \alpha_n(t)]c_{i,k_{n+1}}, \quad \alpha_n = \frac{t_{n+1} - t}{t_{n+1} - t_n},
\end{equation}
yielding a weighted average between consecutive cluster centroids. This approach is posited to offer superior dynamical fidelity for periodic systems compared to CROMs.

Our research introduces a cluster regression model, as detailed in Section 2 of the main manuscript, that captures cluster dynamics without presuming specific transition behaviors between clusters. This model addresses the limitations of CROMs by providing a more flexible framework to model complex system dynamics, potentially enhancing accuracy and applicability in a broader range of scenarios.

\begin{figure}[p]
	\centering
	\includegraphics[width=0.9\textwidth]{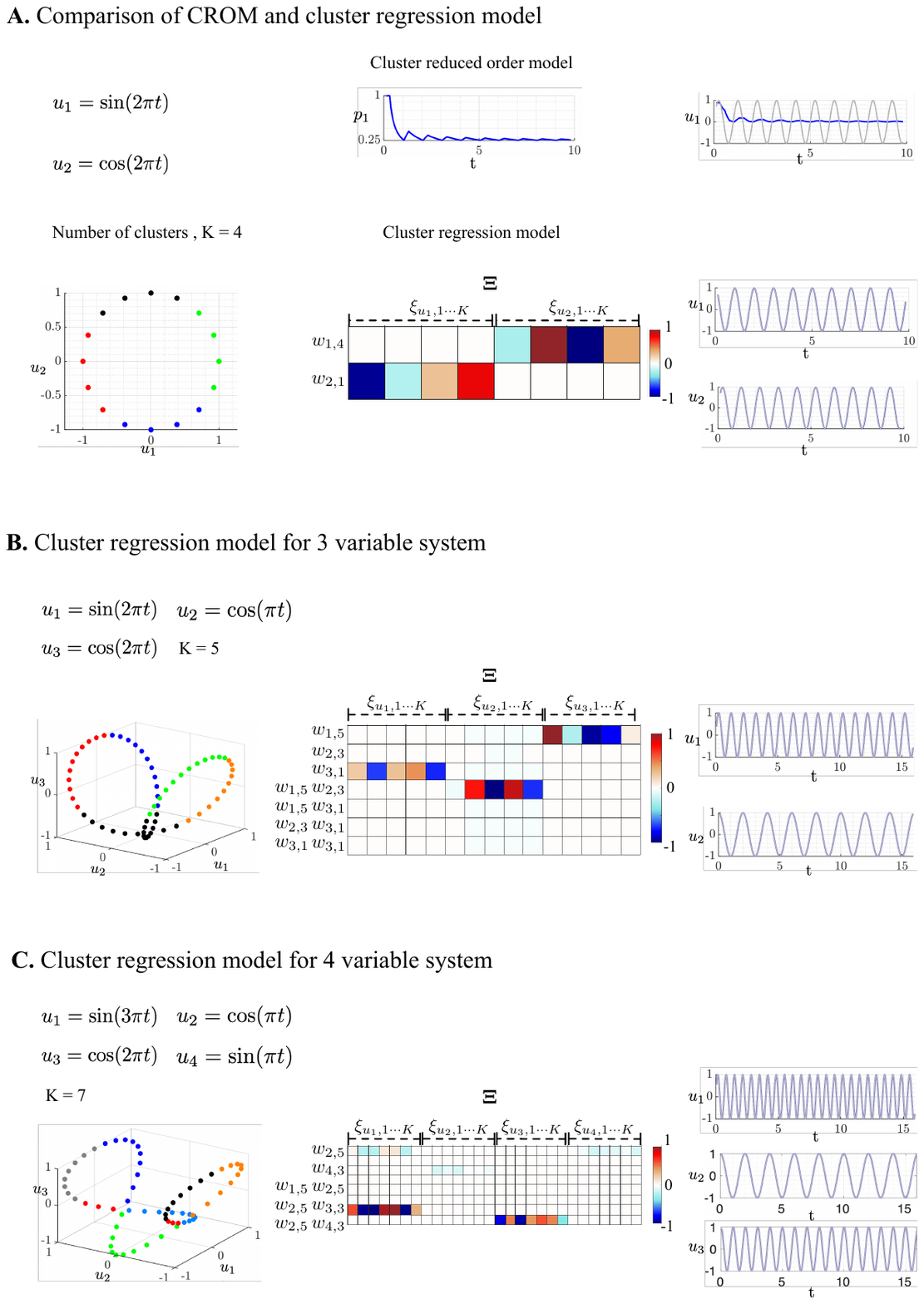}
	\caption{Model assessment for periodic oscillators in multiple dimensions. In each figure, original trajectory is presented in the left panel, middle panel presents the model terms and right panel highlights the model prediction (blue color) compared with the original system (gray color) . \textbf{A}. Comparison of predictive capabilities of cluster regression model and CROM \cite{kaiser2014cluster} for a 2D oscillator system. \textbf{B}. Model construction and prediction for 3 variable system with $5$ clusters. \textbf{C}. Model construction and prediction for 4 variable system with $7$ clusters.}
	\label{fig:A1}
\end{figure}

\section{Model performance for periodic data}
\label{appendixC}

In this section, we evaluate the performance of the cluster regression model for simple periodic systems. Our primary goal is to juxtapose the efficacy of the cluster regression model against the cluster-reduced order model (CROM) for a rudimentary system and then utilize the cluster regression model to forecast periodic systems with multifarious frequencies.

\subsection{Analysis of a single frequency 2D oscillator system}
\label{appendixC1}

Using a basic 2D oscillator, as illustrated in Fig.~\ref{fig:A1}(A) and reminiscent of the model presented by Li et al.~\cite{li2021cluster}, we describe the system's evolution through position vectors at various time points. The system follows the equations \( u_1 = \sin(2 \pi t) \) and \( u_2 = \cos(2 \pi t) \), tracing a unit circle trajectory. Dividing this trajectory into \( K=4 \) clusters, each cluster is color-coded in Fig.~\ref{fig:A1}(A), left panel. With a simulation time step of \( dt=\frac{1}{16} \), transitions between clusters occur within at most three steps, resulting in the trajectory occupying each cluster for approximately a quarter of the total simulation time.

In a Cluster-based Reduced Order Model (CROM), a state probability vector is used to represent system states, which for each state, eventually gravitates towards \( \frac{1}{4} \). This is depicted in the middle panel of Fig.~\ref{fig:A1}(A). Despite starting with higher values, the state probability vector quickly converges to its asymptotic value. However, the rapid dispersion of the state probability vector causes the feature space expectation value \( u_1 \), as calculated by Eq.~\ref{eq:6}, to diverge from the actual dynamics of the oscillator. This limitation undermines the effectiveness of CROM for dynamic system prediction and control.

Conversely, simulations using the cluster regression model (shown in Fig.~\ref{fig:A1}(A), middle panel) capture the system's intrinsic dynamics more accurately. Representing the oscillator with \( \dot{x} = 2\pi \cos(2\pi t) \) and \( \dot{y} = -2\pi \sin(2\pi t) \), starting at \( x=0, y=1 \) at \( t=0 \), we get \( \dot{x} = 2\pi u_2 \) and \( \dot{y} = -2\pi u_1 \). The cluster regression model accurately reflects this relationship through cluster coefficients, as seen in the middle panel of Fig.~\ref{fig:A1}(A). The system displays two synchronized signals with a \( 90^\circ \) phase shift, introducing asymmetry in the model terms. This phase difference contrasts with a scenario of concurrent frequencies without phase differences, which would result in symmetrical model structures. Therefore, the model not only provides comprehensive insights into the system's dynamics but also accurately predicts the dynamics of variables, as shown in the right panel of Fig.~\ref{fig:A1}(A).

\subsection{Analysis of a multi-frequency oscillator system}
\label{appendixC2}
Expanding our analysis, we investigated periodic oscillators in higher-dimensional feature spaces, as shown in Fig.~\ref{fig:A1}(B) for a 3-dimensional system and in Fig.~\ref{fig:A1}(C) for a 4-dimensional system. We set the number of clusters to \( K=5 \) for the 3D oscillator and \( K=7 \) for the 4D oscillator. These cluster numbers, although they could be algorithmically determined, were chosen for clarity based on visually distinct segments of the trajectories.

In contrast to the simple rotational dynamics observed in Fig.~\ref{fig:2}(A), the oscillators in Fig.~\ref{fig:A1}(B) and Fig.~\ref{fig:A1}(C) demonstrate more complex 3D dynamics. Specifically, the oscillator in Fig.~\ref{fig:A1}(B) features dual frequencies, while the one in Fig.~\ref{fig:A1}(C) operates with three different frequencies. In both scenarios, a combination of linear and quadratic terms is employed to effectively describe the coefficient dynamics. 

The cluster regression model identifies $7$ library columns, \( \Theta \), for the regression of the 3-variable system, and $8$ columns for the 4-variable system. This model successfully captures the essential dynamics of the systems. The predictions from the model closely align with the observed multi-frequency dynamics of the original systems, as can be seen in the right panels of Fig.~\ref{fig:2}(B) and Fig.~\ref{fig:A1}(C). The consistency of the model's performance across these examples suggests its potential applicability to complex systems exhibiting diverse temporal scales.

\section{Model and control performance for the Lorenz system}
\label{appendixD}

\subsection{Data generation}
\label{appendixD1}

The Lorenz system's dynamics are evaluated using a fourth-order explicit Runge-Kutta method, initialized with \( x_0 = [-8, 8, 27] \). As referenced in the main manuscript (section 3.1), time series measurements of the variables \(x\), \(y\), and \(z\) are extracted, collectively forming the feature space for the system. Data for clustering is procured from simulations conducted over a duration \( T = 20\, \text{s} \) at a fixed step size \( dt = 0.001\, \text{s} \). The initial \( 10\, \text{s} \) of this data, subject to a phased harmonic Schroeder forcing~\cite{schroeder1970synthesis}, is earmarked for training. In contrast, the latter \( 10\, \text{s} \), modulated by a cubic sinusoidal forcing \( 5\sin(30t)^3 \), is reserved for testing. This division echoes the approach adopted by Kaiser et al.~\cite{kaiser2018sparse}. 

\subsection{Cluster resolution}
\label{appendixD2}

\begin{figure}
	\centering
	\includegraphics[width=1.00\textwidth]{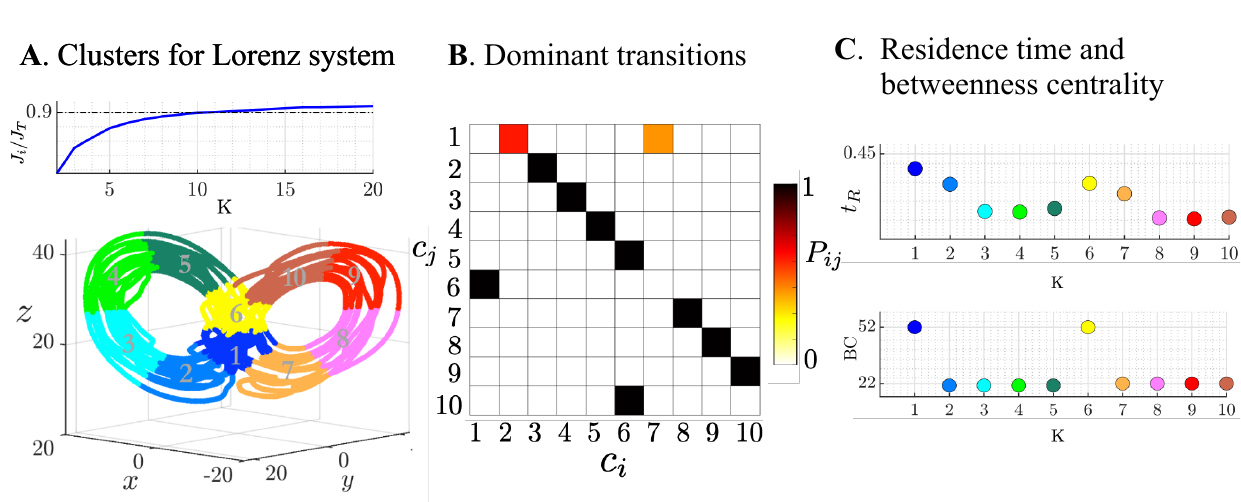}
	\caption{Cluster analysis of the forced Lorenz system. \textbf{A}. (top panel) The optimum number of clusters ($K=10$) found using F-test \cite{lomax2013introduction} criterion. Visualization of the clusters for the Lorenz attractor. \textbf{B}. The dominant transitions for the system. Each element represents transition probability from cluster $c_j$ to $c_i$. \textbf{C}. (top panel) Measure of time spent in a particular cluster called as the cluster residence time. (bottom panel) Measure of connectivity of a cluster with other clusters called as betweenness centrality. }
	\label{fig:A2}
\end{figure}

To determine the appropriate cluster resolution, we deploy an F-test criterion. This identifies \( K = 10 \) as the optimal cluster number, as showcased in the upper panel of Fig.~\ref{fig:A2}(A). The Lorenz attractor's clusters, vividly color-coded, are depicted in Fig.~\ref{fig:A2}(A)'s bottom panel. Upon closer scrutiny, the clusters organically separate into three discernible groups. Clusters 2, 3, 4, and 5 delineate the attractor's left lobe, whereas clusters 7, 8, 9, and 10 define its right counterpart. Clusters 1 and 6 serve as the bridge between these lobes. Such pivotal clusters are designated as flipper clusters~\cite{kaiser2014cluster} due to their facilitative role in lobe transitioning.

\subsection{Transition matrix}
\label{appendixD3}
The computed transition matrix, directly derived from the data, is illustrated in Fig.~\ref{fig:A2}(B). An element \( P_{ij} \) of this matrix signifies the transition from cluster \( c_j \) to cluster \( c_i \). This matrix unveils intriguing insights into the trajectory's flow, which can be harnessed for control purposes. For instance, trajectories from cluster 2 invariably transition to cluster 3, and subsequently, from cluster 3 to cluster 4. The trajectory within flipper cluster 1, on the other hand, can shift to either cluster 2 or cluster 7. In the context of the Lorenz system, the flipper cluster can generate two states: one with the trajectory ensconced in the left lobe, and the other in the right lobe. Recognizing the flipper cluster is paramount as it suggests where conditional control can be applied to regulate the trajectory. To preclude lobe switching, controls could, for instance, be enforced on cluster 1. It merits emphasis that the transition matrix is leveraged solely for analytical purposes, not state propagation (which is the purview of the predictive model). However, this transitional data remains invaluable for both analytical assessments and the design of control strategies.

In the Lorenz system, the transition behavior elucidates the trajectory's flow. To elucidate, a trajectory in cluster 2 can only transition to cluster 1 via the sequence: \(2 \rightarrow 3 \rightarrow 4 \rightarrow 5 \rightarrow 6 \rightarrow 1\). While this information might appear elementary for the Lorenz system, it offers insights into various physical processes and the underlying mechanisms in high-dimensional systems~\cite{kaiser2014cluster, nair2019cluster, barwey2023data}. Even though the cluster regression model capitalizes on constrained sensor measurement data, the high-dimensional physical processes mirrored in the snapshots can be discerned through their cluster affiliations. This approach diverges significantly from the clustering of POD coefficients ~\cite{kaiser2014cluster} and the clustering of the complete snapshot datasets ~\cite{barwey2023data}.

\subsection{Cluster importance}
\label{appendixD4}

The cluster residence time, denoted by \( t_R \), signifies the fraction of the entire time the trajectory resides within a specific cluster. This metric is depicted in the top panel of Fig.~\ref{fig:A2}(C). The predominant residence time is attributed to cluster number $1$, succeeded by clusters $6$ and $2$. The central positioning of clusters $1$ and $6$—serving as conduits between the two lobes—necessitates the trajectory's passage through them. Additionally, the substantial residence time in cluster $2$ supersedes that of other clusters, attributed to the attractor's structure, which amplifies the span of cluster $2$.

For high-dimensional systems, the cluster residence time unveils pivotal insights regarding states where the system predominantly or scarcely lingers. Furthermore, this time can serve as an analogue for stability analysis in nonlinear systems~\cite{lakshmikantham1989stability}. Clusters with diminished residence times might be perceived as states where the system exhibits relative instability, thus being more susceptible to perturbations. Conclusively, the cluster residence time can be instrumental in devising closed-loop control strategies by pinpointing the clusters that are acutely responsive to control.

The concept of betweenness centrality measures the centrality of a node or group, in our context, a cluster, in a network. This metric accentuates nodes that act as conduits for the propagation of information throughout a network \cite{brandes2001faster,rubinov2010complex}. This is illustrated in the lower panel of Fig.~\ref{fig:A2}(C), where clusters $1$ and $6$ prominently exhibit the highest betweenness centrality values. Notably, these clusters also correspond to those with the most prolonged residence times. While this correlation is conspicuous for the Lorenz system, it might not be universally consistent. The betweenness centrality proves invaluable for pinpointing the flipper cluster. Within a network of clusters, a cluster interconnected with the majority of other clusters plays a pivotal role in information dissemination. Exerting control on such a cluster can induce substantial modifications in the system's characteristics.

\subsection{Model predictive control (MPC) setup}
\label{appendixD5}

\begin{figure}
	\centering
	\includegraphics[width=0.9\textwidth]{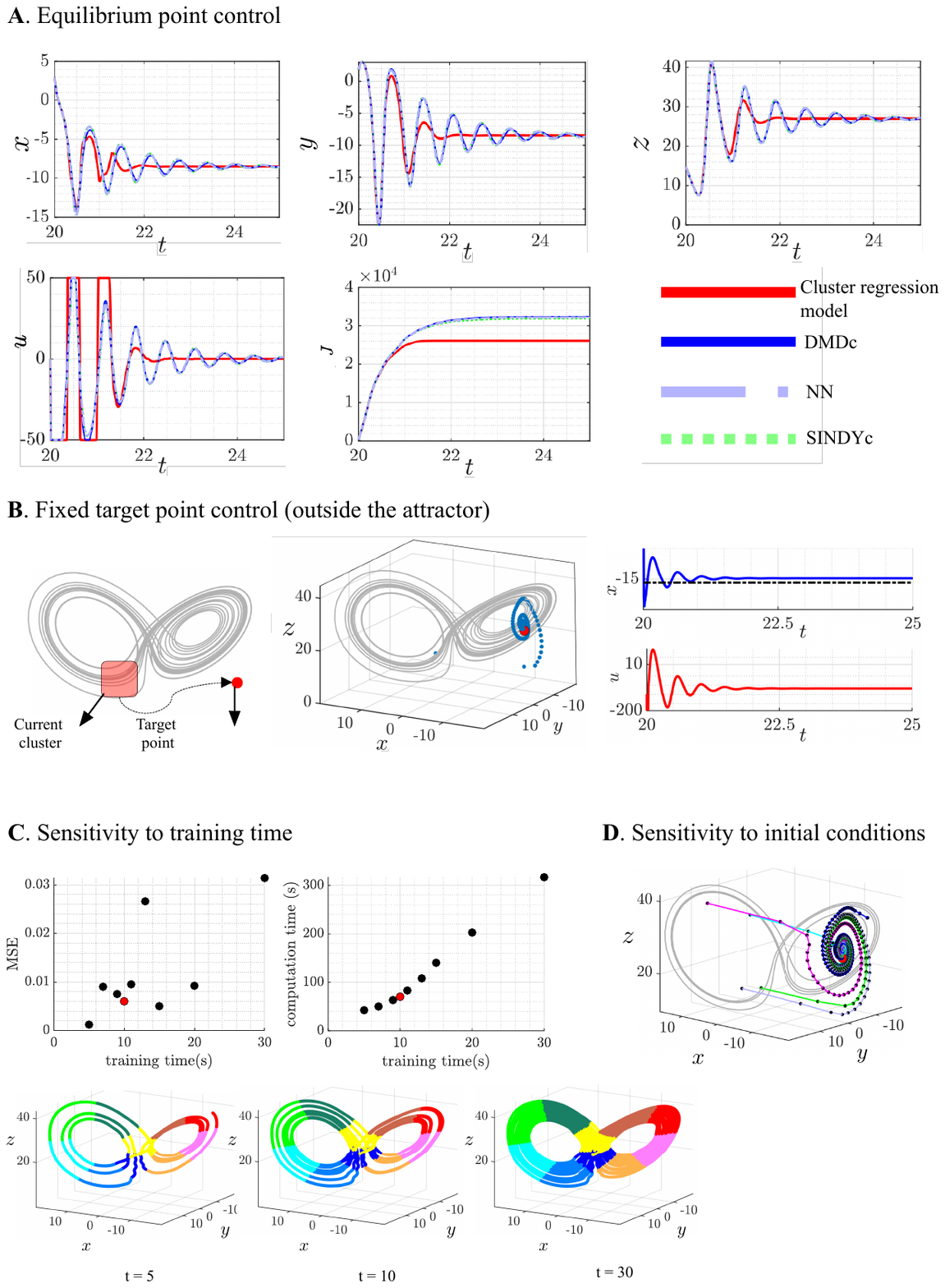}
	\caption{Analysis of Lorenz system with control. \textbf{A}. Comparison between the cluster regression model and dynamic mode decomposition with control, sparse identification of nonlinear dynamics with control and neural network model with control \cite{kaiser2018sparse}. \textbf{B}. Model performance for control of a point lying outside the attractor. \textbf{C}. Sensitivity of the model to training time. Top panel shows the mean squared error and physical computation time against training time of the model. Bottom panel shows the clusters for Lorenz attractor with training times. The red dot represents the choice of the training time for results reported in this work. \textbf{D}. Control of fixed point with different initial conditions.}
	\label{fig:A3}
\end{figure}

The cluster regression model employs a $\textup{2nd}$ order polynomial regression with the QR-truncated library to formulate the dynamic model, setting $\lambda = 0$. Subsequently, this model is integrated with MPC to regulate the chaotic dynamics. Our primary control objective targets stabilizing the equilibrium fixed point at $(-\sqrt{72},-\sqrt{72},27)$. Post the training and validation over a $20s$ duration, the Lorenz system's control via MPC is executed for $5s$, with both the prediction and control horizons set at $10$ steps. The weight matrix is denoted as $Q=I_3$ (with $I_n$ representing an $n\times n$ identity matrix). The controller cost weight matrices are given by $R=0.001$ and $R_{\Delta u} = 0.001$, while the actuation input is constrained to $u \in [-50,50]$. For equilibrium point control in the Lorenz system, we juxtapose the cluster regression model with established models, such as SINDYc, DMDc, and a neural network model \cite{kaiser2018sparse}, as depicted in Fig.~\ref{fig:A3}(A).

In the application of the cluster regression model, the trajectories of $x$, $y$, and $z$ converge to the equilibrium point more rapidly compared to alternative models. Initially, the control input for this model fluctuates between extremal values, exceeding the actuation magnitude of its counterparts. Yet, its prompt steering of the trajectories towards the equilibrium diminishes the need for subsequent control interventions. Consequently, the aggregate cost of control for the cluster regression model is reduced. Overall, the cluster regression model demonstrates superior performance in guiding the system to its equilibrium state.

\subsection{Model and control performance}
\label{appendixD6}

The model's predictive capability is validated against training and testing data, as well as for controlling the system to an equilibrium state. We extend our evaluation to control objectives that are neither part of the training nor the test datasets, nor represent an equilibrium point. Figure \ref{fig:A3}(B) illustrates our attempt to direct the system from an initial cluster (light red box) to a target point $(-15.58,-4.42,25.41)$ located off the attractor (dark red point in the left panel of Fig. \ref{fig:A3}(B)). Achieving this control requires significant actuation, potentially increasing computational costs and inducing oscillatory behavior in the trajectory. However, the chosen target point does not necessitate ad-hoc control adjustments to maintain tractable solutions. For this task, we adjust the control parameters to $R=0.0001$ and $R_{\Delta u} = 0.0001$, which are sufficient to guide the trajectory to the target effectively.

This capability is crucial in scenarios like unsteady fluid flow, where control outside the range of known dynamics—such as mitigating lift transients from an unexpected gust—adds considerable value. As shown in the right panel of Fig. \ref{fig:4}(B), a substantial initial actuation is required to displace the trajectory from the attractor. Once near the target, the control effort is reduced. 

\subsection{Model sensitivity}
\label{appendixD7}

We assess the sensitivity of the cluster regression model to the duration of the training period, as illustrated in Fig. \ref{fig:A3}(C). The model's performance, measured by the mean squared error (MSE) during the testing phase, is compared across different training durations. Notably, the computation times for both the training and testing phases are documented, with the case employed in our study highlighted by red dots. The analysis reveals that the MSE attains its minimum for a training duration of $t=5$ seconds and peaks at $t=30$ seconds. This trend is expected since shorter training durations correspond to more concise testing periods, inherently yielding lower prediction errors due to the less complex dynamics captured within a developing attractor. However, a shorter training period, like $t=5$ seconds, results in a poorly resolved cluster distribution, which may lead to flawed interpretations in cluster analysis due to under-represented states.

In contrast, for a training period of $t=10$ seconds, the MSE remains within acceptable bounds, and the attractor exhibits a well-distributed, mature cluster structure. Additionally, the computational demand for this training duration is significantly lower than that for $t=30$ seconds, striking an optimal balance between accuracy and efficiency. Consequently, we have established $t=10$ seconds as the standard training duration for our cluster regression model.

To assess the control sensitivity for different initial conditions,  we choose few points scattered along different clusters over the attractor (only four of these points are shown in Fig. \ref{fig:A3}(D) for clarity ) with target point outside the training and test data.  It is observed from Fig. \ref{fig:A3}(D)  that the trajectories from various clusters are successfully steered to the off-attractor target, each time withdrawing from the attractor before settling into the target region.

\section{Flow over a compliant flat plate}
\label{appendixE}

\subsection{Data generation}
\label{appendixE1}

We employ a cluster regression model with control for a DNS of an incompressible flow over a compliant thin plate, positioned at an angle of attack $\alpha = 35^{\circ}$, using the immersed boundary method \cite{taira2007immersed,goza2017strongly}. This configuration induces separation at both the leading and trailing edges, leading to periodic vortex shedding and fluctuating aerodynamic forces. The computational process is expedited using a multi-domain technique \cite{colonius2008fast}. Within this immersed boundary framework, the fluid flow is governed by:

\begin{equation}
	\nabla \cdot \mathbf{u} = 0
\end{equation}

\begin{equation} \label{eq:16}
	\frac{\partial \mathbf{u}}{\partial t} + (\mathbf{u} \cdot \nabla)\mathbf{u}  = -\nabla p + \frac{1}{\text{Re}}\nabla^2 \mathbf{u} + \int_\Gamma \mathbf{f}(\bm{\chi}(s,t))\delta(\bm{\chi}(s,t)-\mathbf{x})\mathrm{d}s
\end{equation}

\begin{equation}
	\frac{\rho_s}{\rho_f}\frac{\partial^2 \bm{\chi}}{\partial t^2} = \frac{1}{\rho_fU^2}\nabla \cdot \bm{\sigma} + \mathbf{g}(\bm{\chi}) - \mathbf{f}(\bm{\chi})
\end{equation}

\begin{equation}
	\int_\Gamma \bm{u}(\mathbf{x})\delta(\mathbf{x} - \bm{\chi}(s,t))\mathrm{d}\mathbf{x} = \frac{\partial \bm{\chi}(s,t)}{\partial t}
\end{equation}

Here, $\bm{\chi}$ indicates the Lagrangian co-ordinate attached to the flexible body $\Gamma$, $\mathbf{x}$ is the Eulerian co-ordinate and $s$ is the co-ordinate that parametrizes the body shape. $\mathbf{u}$ is the flow velocity, while $\rho_f$ and $\rho_s$ denote fluid and solid densities, respectively. $\bm{\sigma}$ represents the Cauchy stress and $\mathbf{f}(\bm{\chi})$ denotes the surface stress imposed on the
fluid by the immersed body.  Convective terms are discretized through the Adams-Bashforth method, and viscous terms via the implicit Crank-Nicolson scheme. Large displacements in the solid are addressed using an Euler-Bernoulli beam equation, combined with corotational finite element discretization.

The flat plate is anchored at its leading edge ($x/c=0,y/c=0$) and is discretized into 65 segments. The non-dimensional parameters governing this fluid-structure interaction configuration are the bending stiffness $K_B = \frac{EI}{\rho_f^2U^3_{\infty}c^3}$, mass ratio $M_{\rho} = \frac{\rho_sh}{\rho_fc}$, and Reynolds number $R_e = U_{\infty}c/\nu$. Our setup employs a plate with flexibility parameter $K_B = 0.3125$, mass ratio $M_\rho = 3$, and Reynolds number $R_e=100$, similar to previous work by Hickner et al. \cite{hickner2023data}. The simulation utilizes five grid levels, with the innermost domain as $ -0.2 \leq x \leq 1.8$ and $ -1 \leq y \leq 1$, and maintains consistent grid spacing of $\Delta x/c \approx 0.0077$.

We consider the time-evolving lift coefficient and its derivative, along with the drag coefficient, to form the feature space for clustering of this high-dimensional system. The section lift and drag coefficients are computed using the relations:
\begin{equation}
	C_D = \frac{2D}{\rho_{\infty}U_{\infty}^2c}, \quad C_L = \frac{2L}{\rho_{\infty}U_{\infty}^2c}
\end{equation}
where $D$ and $L$ represent the drag and lift forces, respectively, $\rho_{\infty}$ is the free-stream density, $U_{\infty}$ the free-stream velocity, and $c$ the chord length of the plate. The feature space of $[C_L(t), \dot{C}_L(t), C_D(t)]$ capture the unsteady behavior that the system experiences \cite{nair2019cluster,noack2003hierarchy,taira2018phase}.

The training of our model incorporates forcing, analogous to the procedure with the Lorenz system. We employ phased harmonic Schroeder forcing through a momentum jet strategically positioned at the plate's leading edge. The forcing mechanism mimics a suction and blowing actuator and is introduced into the momentum equation (see Eq. \ref{eq:16}) as a source term. The orientation of the actuation is normal to the oncoming flow, effectively modulating the circulation over the airfoil and generating a diverse array of lift and drag states for model training.

The effectiveness of the actuation is quantified by the momentum coefficient, given by:
\begin{equation}
	\bar{C}_\mu = \frac{2\sigma}{TU_{\infty}^2c} \int_0^{T} |u_{jet}|^2 \, \mathrm{d}t
\end{equation}
In this formula, \( |u_{jet}|^2 \) represents the squared magnitude of the jet velocity, \( \sigma \) the width of the actuation slot, and \( T \) the duration of actuation. As our setup lacks a physical slot, we simulate the actuation using a regularized delta function with an effective width (\(\sigma\)) that matches the spatial discretization step (\(\Delta x\)), yielding a ratio \(\sigma/c \approx 0.0077\). To minimize numerical errors that could arise from the proximity of the actuator to the boundary forces, the momentum is injected at a vertical distance of \( 8\Delta x/c \) above the leading edge \cite{roma1999adaptive, taira2010lift}

\begin{figure}
	\centering
	\includegraphics[width=1.00\textwidth]{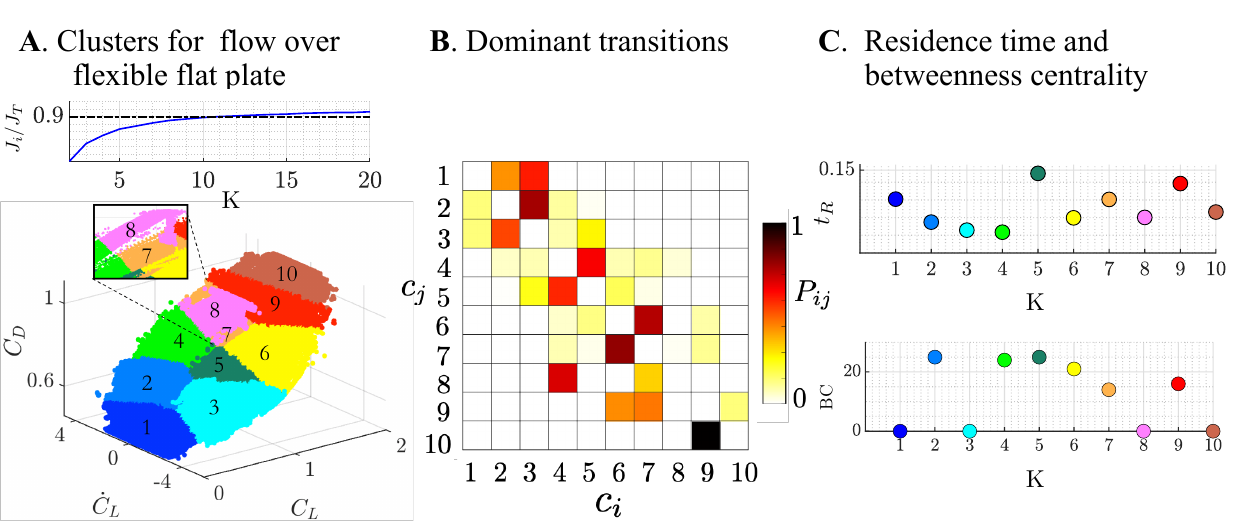}
	\caption{Cluster analysis for flow over flexible flat plate at $Re=100$. \textbf{A}. (top panel) The optimum number of clusters ($K=10$) found using F-test criterion. (bottom panel) Visualization of the clusters. \textbf{B}. The dominant transitions for the system. Each element represents transition probability from cluster $c_j$ to $c_i$. \textbf{C}. (top panel) The cluster residence time and (bottom panel) betweenness centrality highlighting the cluster importance. }
	\label{fig:A4}
\end{figure}

\subsection{Cluster resolution}
\label{appendixE2}

The dataset from the DNS extends over ten convective time units, each corresponding to a full cycle of vortex shedding in the baseline case. The selection of the optimal number of clusters is informed by an F-test, the results of which are displayed in the upper panel of Fig.~\ref{fig:A4}(A), suggesting \( K=10 \) as the ideal cluster count. The identified clusters are visually represented in the lower panel of Fig.~\ref{fig:A4}(A), providing a clear conceptual map of the flow's lift and drag states. It is particularly noted that higher lift coefficients are often associated with higher drag coefficients. A cluster regression model is then devised to forecast the lift (\( C_L \)) and drag (\( C_D \)) coefficients, applying a first-order polynomial regression approach.

\subsection{Transition matrix}
\label{appendixE3}

The transitions between clusters, akin to those observed in Fig.~\ref{fig:A2}(B), are delineated in Fig.~\ref{fig:A4}(B). The system exhibits a range of lift and drag states as a result of the interplay between the vortex shedding dynamics and the effects of jet injection, which modulates the circulation around the plate. Notably, cluster $10$ is associated with the peak values of lift and drag and is most commonly reached via transitions from cluster $9$, albeit with low probability, indicating that transitions to states of higher lift are energetically less favorable. Transitions involving cluster $8$ are particularly interesting; it serves as an intermediate state, predominantly transitioning to cluster $7$ or reverting to cluster $4$, suggesting a transient increase in drag followed by a rapid decrease, potentially attributable to the dynamics of vortex shedding or flow elongation. The state of minimal drag is represented by cluster $1$, which typically transitions to clusters $2$ or $3$. The structure of the transition matrix implies a modicum of compartmentalization among the clusters, a hypothesis further substantiated by a community detection algorithm that identifies three distinct subgroups within the transition matrix: clusters $1$, $2$, and $3$ form one group; clusters $4$, $5$, $6$, and $7$ another; and clusters $8$, $9$, and $10$ the final one. Subsequent sections will demonstrate how targeted control actuations can influence these predominant transition pathways, effectuating changes in the system's dynamic states.

\subsection{Cluster importance}
\label{appendixE4}

The distribution of cluster residence times, depicted in Fig.~\ref{fig:A4}(C), indicates that cluster $5$ exhibits the longest duration of persistence within the system. Similar to the observations made for the Lorenz system, the cluster occupying a central position in the state space generally manifests a prolonged residence time and is characterized by a high betweenness centrality metric. In the current analysis, clusters $5$, $2$, and $4$ register the highest betweenness centrality values, underscoring their pivotal roles in the state transition network. The existence of multiple clusters with flipper-like behavior introduces a layer of complexity to the control strategy. Specifically, pinpointing the most effective flipper cluster for control purposes is a nontrivial task, which complicates efforts to harness the system's inherent dynamics for managing the flow regime.

\subsection{Model predictive control (MPC) setup}
\label{appendixE5}

The integration of MPC with DNS is implemented for the actuation mechanism. The MPC employs a prediction and control horizon of $10$ steps each. The quadratic penalty on the state deviation is captured by the weight matrix $Q=I_3$, where $I_n$ denotes an $n \times n$ identity matrix. The controller cost weight matrices are set with $R=0.001$ and $R_{\Delta u}=0.001$, reflecting a moderate penalty on control effort and changes in control input, respectively. The actuation velocity, determined by the MPC, is applied in the DNS as boundary conditions, with its maximum value constrained to avoid numerical instability within the simulation framework.

\begin{figure}
	\centering
	\includegraphics[width=0.9\textwidth]{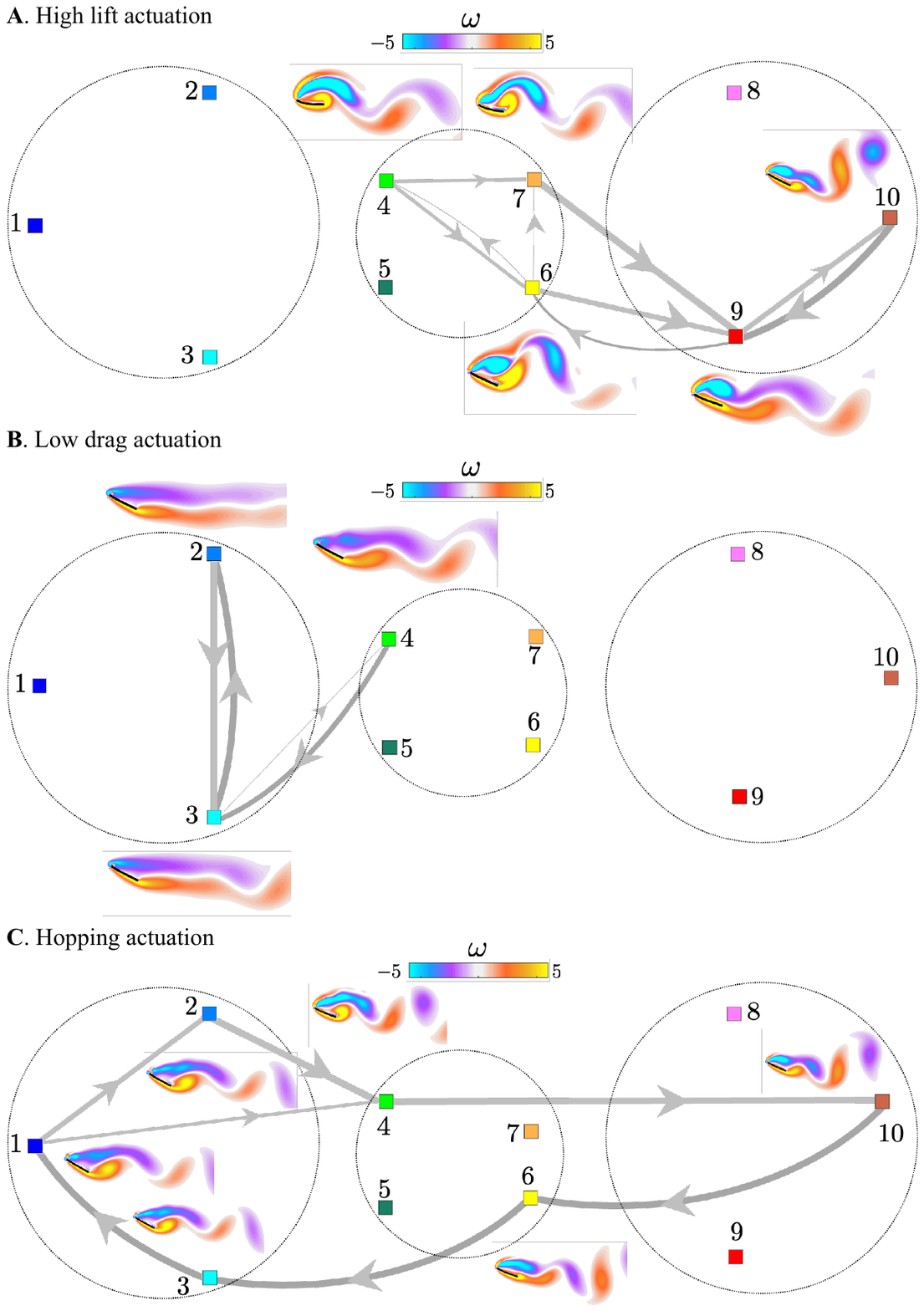}
	\caption{Dominant transitions in flow control. The thickness of the arrows represent the transition probability. \textbf{A}. High lift actuation: The flow resides in high lift clusters and medium lift clusters. \textbf{B}. Low drag actuation: The flow predominantly resides in low drag clusters. \textbf{C}. Hopping actuation: The flow hops between high lift and low drag clusters.}
	\label{fig:A5}
\end{figure}

\subsection{Dominant transitions in flow control}
\label{appendixE6}

Figure~\ref{fig:A5} elucidates the predominant transition dynamics under various flow control scenarios, with transition probabilities symbolized by arrow thickness. Additionally, we provide vorticity contour visualizations to illustrate the complex, high-dimensional flow field shaped by the control interventions.

Figure~\ref{fig:A5}(A) showcases the flow topology linked to aggressive lift enhancement, targeting cluster $10$. Here, the transitions skew heavily towards clusters associated with elevated lift. Clusters $9$ and $10$ display considerable circulation above the plate, which underpins the intensified lift. Transitions to clusters with augmented lift are markedly increased, while shifts to lower drag clusters ($1$, $2$, and $3$) are conspicuously absent. Clusters $4$ through $7$, embodying medium lift states, correspond to flow configurations where a decline in lift is imminent, instigated by interactions between leading and trailing edge vortices or through vortex shedding. Upon a decrease in lift, the controller promptly intervenes, injecting momentum to bolster the lift, thereby nudging the system back towards higher lift states. 

Conversely, Figure~\ref{fig:A5}(B) captures the flow's bias towards a low drag regime, specifically cluster $2$. A marked reduction is noted in transitions towards higher lift/higher drag clusters, with the flow largely settling in lower drag states. The most prevalent transitions are between clusters $2$ and $3$. The vorticity contours depict stretched vortices indicative of low drag conditions.

\begin{figure}
	\centering
	\includegraphics[width=0.9\textwidth]{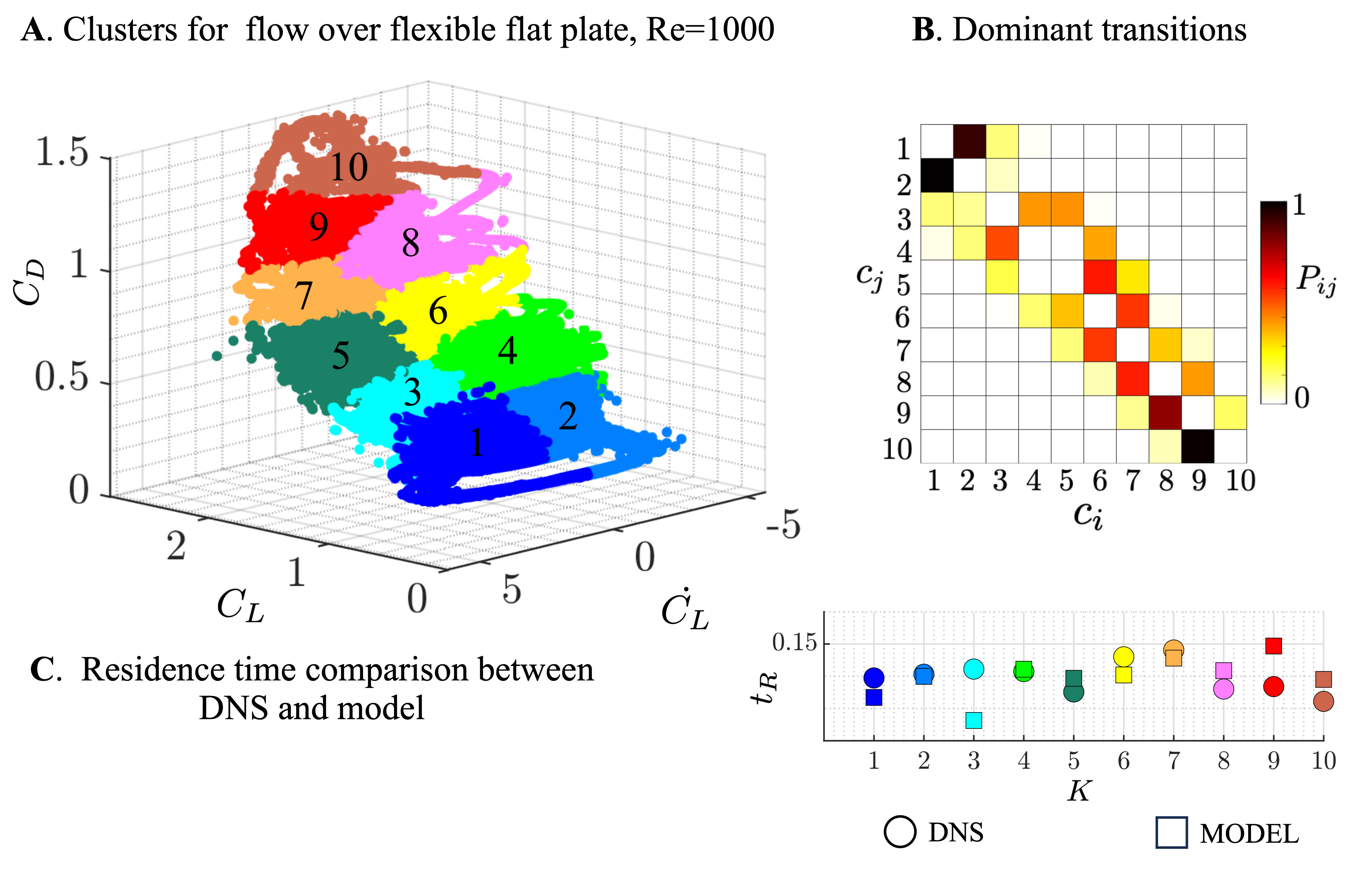}
	\caption{Flow over flexible flat plate at $Re=1000$. \textbf{A}. The trajectory divided into $K=10$ clusters. \textbf{B}. The dominant transitions for the system. Each element represents transition probability from cluster $c_j$ to $c_i$. \textbf{C}. Comparison of the cluster residence time obtained though DNS data and the cluster regression model.}
	\label{fig:A6}
\end{figure}

Across both scenarios, cluster $4$ emerges as pivotal in the control strategy. When guiding the flow towards high lift, transitions from low lift clusters to cluster $4$ are suppressed; when steering towards low drag, transitions from cluster $4$ to higher drag clusters are curtailed. Cluster $4$, previously identified as a 'flipper' cluster, thus appears to have its dynamics substantially altered during control. 

In the 'hopping' actuation scenario shown in Figure~\ref{fig:A5}(C), cluster $4$ is instrumental in facilitating transitions from low drag to high lift states, with a direct path from cluster $4$ to $10$ becoming apparent. Transitions from high lift to low drag, however, proceed through cluster $6$-another flipper cluster-directly to cluster $3$. The transitions in Fig. ~\ref{fig:A5}(C) hint at a hysteresis loop in the system, congruent with the idea that high lift states are induced by momentum injection, while low drag states arise from disparate phenomena like vortex stretching, shedding, or mutual interaction of vortical structures. The identification of a hysteresis loop corroborates the existence of multiple flipper clusters, aligning with earlier insights.

\begin{figure}[h!]
	\centering
	\includegraphics[width=0.85\textwidth]{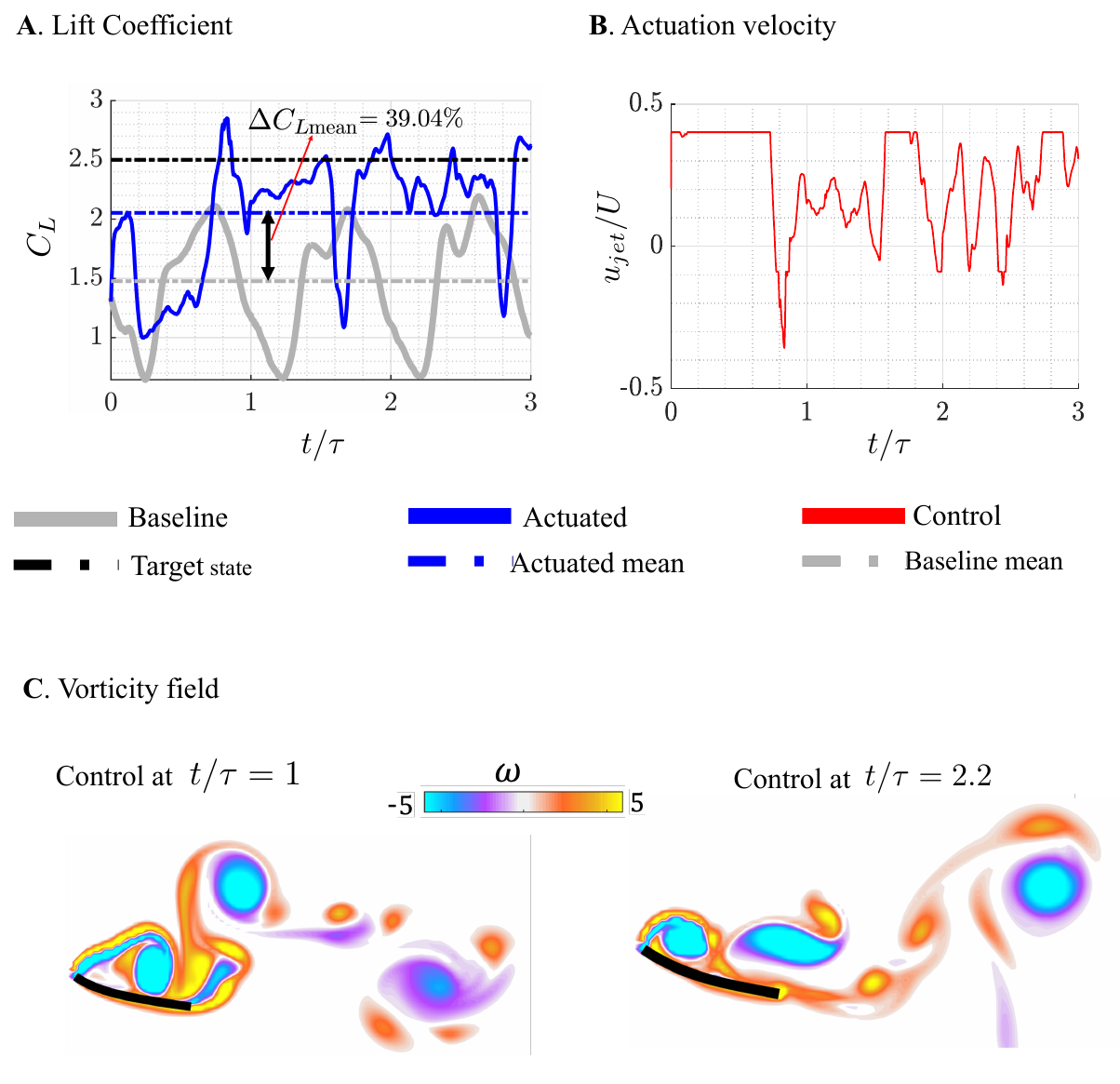}
	\caption{High lift actuation for $Re=1000$. \textbf{A}. Comparison of the lift coefficient resulting from actuation with the baseline. \textbf{B}. The actuation velocity during the control sequence. \textbf{C} The resulting vorticity field displaying high circulation over the plate.}
	\label{fig:A7}
\end{figure}

\subsection{Model and control performance at higher Reynolds number}
\label{appendixE7}

We explore the control towards a high lift state at a higher Reynolds number, $Re = 1000$. The fundamental behavior at $Re = 100$ is defined by periodic limit cycle oscillations in the force coefficients, whereas at $Re = 1000$, we observe a quasi-periodic regime. The interaction between the trailing edge and leading-edge vortices at this higher Reynolds number introduces nonlinearities at the trailing edge of the plate, thereby affecting the cycle-to-cycle vortex shedding response, as detailed by Liu \textit{et al.}~\cite{liu2021aerodynamic} and Williamson~\cite{williamson1988vortex}. The complex vortex dynamics at play pose a greater challenge for the model, especially when flexibility and additional forcing are introduced for training purposes.

Following the established training methodology, we delineate the resulting clusters in Fig.~\ref{fig:A6}(A), observing a segmentation of the dataset into discrete lift and drag states. Contrary to the $Re=100$ scenario, dominant transitions depicted in Fig.~\ref{fig:A6}(B) tend to aggregate near the diagonal, indicating a reduced extent of cluster interaction. Furthermore, the distribution of cluster residence times, as shown in Fig.~\ref{fig:A6}(C), exhibits minimal variability, hinting at a constricted interchange between low drag and high lift states and a diminished sensitivity to the applied forcing. Comparative analysis of cluster residence times derived from DNS showcases a respectable concordance with those procured from the model predictions.

The execution of control towards a high lift state for $Re = 1000$ is graphically represented in Fig.~\ref{fig:A7}. All control parameters are retained from the $Re=100$ setup. During the control application, the lift coefficient trajectory reveals several notable downturns, attributed to the formation and shedding of sizeable vortices as a result of jet injection. Besides these prominent fluctuations, smaller downturns are also evident, arising when the leading edge vortex deviates from the plate, an effect of either plate deflection or trailing edge vortex influence. This complex dynamic is reflected in the jet actuation velocity profile, which displays a series of peaks. 

The resulting flow characteristics for the control actuation in this case is significantly different from that for $Re=100$ case. This is partly attributed to the different shedding behavior for the two Reynolds number \cite{williamson1988vortex}. The interaction of control actuation with the flow gives rise to a range of shed vortices visible in the wake which is responsible for the highly irregular actuated lift coefficient profile in Fig. \ref{fig:A7}(A). This is in stark difference to the smooth profile for actuated lift coefficient for $Re=100$ case (left panel of the first row of Fig. \ref{fig:5} in the main manuscript). The existence of minor deflection and bending of the plate in the present case suggests slower lift increments as compared to $Re=100$ case (left panel of the last row of Fig. \ref{fig:5} in the main manuscript) associated with large deflection and bending of the plate. The difference is due to capping of the maximum actuation velocity in the present case which is restricted to nearly $40\%$ of the incoming flow velocity for numerical stability. The actuation velocity and incoming flow velocity were nearly the same order for $Re=100$ case. However, the high lift actuation still yields an approximate $39\%$ enhancement in the average lift coefficient.

\end{appendix}

\bibliographystyle{unsrt}
\bibliography{biblio}
\end{document}